\documentclass[twocolumn,preprintnumbers,amsmath,amssymb,nofootinbib,pra]{revtex4}
\usepackage{graphicx}
\usepackage[applemac]{inputenc}
\usepackage{amsmath}

\begin{document}

\title{Search for non-Gaussianities in the WMAP data with the Scaling Index Method}
			
\author{G. Rossmanith$^1$, H. Modest$^1$, C. R\"ath$^1$, A. J. Banday$^{2,3,4}$, K. M. G\'{o}rski$^{5,6}$ and G. Morfill$^1$}
\affiliation{	$^1$ Max-Planck-Institut f\"ur extraterrestrische Physik, Giessenbachstr. 1, 85748 Garching, Germany \\
		$^2$ Universit\'{e} de Toulouse; UPS-OMP; IRAP; Toulouse, France \\
		$^3$ Centre d’Etude Spatiale des Rayonnements, 9, Av du Colonel Roche, 31028 Toulouse, France \\
		$^4$ Max-Planck-Institut f\"ur Astrophysik, Karl-Schwarzschild-Str. 1, 85741 Garching, Germany \\
		$^5$ Jet Propulsion Laboratory, California Institute of Technology, Pasadena, CA 91109, USA \\
                $^6$ Warsaw University Observatory, Aleje Ujazdowskie 4, 00 - 478 Warszawa, Poland
}

\date{\today}

\begin{abstract}
In the recent years, non-Gaussianity and statistical isotropy of the Cosmic Microwave Background (CMB) was investigated with various statistical measures, first and foremost by means of the measurements of the WMAP satellite. In this \textit{Review}, we focus on the analyses that were accomplished with a measure of local type, the so-called \textit{Scaling Index Method} (SIM). The SIM is able to detect structural characteristics of a given data set, and has proven to be highly valuable in CMB analysis. It was used for comparing the data set with simulations as well as surrogates, which are full sky maps generated by randomisation of previously selected features of the original map. During these investigations, strong evidence for non-Gaussianities as well as asymmetries and local features could be detected. In combination with the surrogates approach, the SIM detected the highest significances for non-Gaussianity to date.
\end{abstract}

\maketitle

\label{firstpage}


\section{Introduction}

The inflationary phase, first proposed in 1981, is an important part of what is called the Standard Model of Cosmology. Since inflation occurred already a few moments after the Big Bang, where the Universe was extremely hot, dense and thus opaque, it is not possible to observe this short time period directly. The best way to achieve information about and to test theories for inflation, is to look at the temperature anisotropies of the Cosmic Microwave Background (CMB). While the simplest single field slow roll inflationary scenario predicts these fluctuations to be nearly Gaussian \cite{1981PhRvD..23..347G, 1982PhLB..108..389L, 1982PhRvL..48.1220A}, a variety of more complex models could lead to a different result (e.g. \cite{1997PhRvD..56..535L, 1997ApJ...483L...1P, 2002PhRvD..66j3506B, 2003NuPhB.667..119A}). By testing the Gaussianity of the CMB, it is possible to distinguish between different inflationary models, and therefore to shed some light on the physics of the very early Universe.

In recent years, the observations of the CMB accomplished by the WMAP satellite offered researchers the possibility to analyse high resolution full sky data maps of this relic radiation. A multiplicity of analyses addressed the challenge of non-Gaussianity, thereby applying several different statistical measures as for example (but far from complete) the angular bispectrum \cite{2003ApJS..148..119K, 2011ApJS..192...18K}, Minkowski functionals \cite{2008MNRAS.389.1439H, 2004ApJ...612...64E}, multipole vectors \cite{2004PhRvL..93v1301S, 2011APh....34..591S}, phase mapping techniques \cite{2003ApJ...590L..65C, 2005PhRvD..72f3512N, 2007ApJ...664....8C, 2007MNRAS.380L..71C}, wavelets \cite{2004ApJ...609...22V, 2007ApJ...655...11C, 2011MNRAS.412.1038C}, needlets \cite{2008PhRvD..78j3504P, 2009MNRAS.396.1682P} or the Kolmogorov stochasticity parameter \cite{2008A&A...492L..33G, 2009A&A...497..343G}. In the course of these investigations, several deviations from Gaussianity and statistical isotropy could be detected, in particular unexpected multipole alignments \cite{2003PhRvD..68l3523T, 2004PhRvD..69f3516D, 2007PhRvD..75b3507C, 2006MNRAS.367...79C}, asymmetries \cite{2004ApJ...605...14E, 2010ApJ...722..452G} or local features - first and foremost the famous \textit{Cold Spot} \cite{2004ApJ...609...22V}.

In this \textit{Review}, we focus on another measure for non-Gaussianity analyses, namely the (weighted) \textit{scaling index method} (SIM) \cite{2002MNRAS.337..413R, 2003MNRAS.344..115R}. The SIM is able to distinguish different structural behaviour of a data set in a \textit{local} way. Scaling indices have already been used for texture discrimination \cite{Rath:97} and feature extraction \cite{Jamitzky01, 1367-2630-10-12-125010}, time series analysis of stock exchanges \cite{Monetti04} and active galactic nuclei \cite{2002A&A...391..875G, 2006A&A...449..969G}, as well as structure analysis of bone images \cite{springerlink:10.1007/s00198-006-0130-1} and other different medical data, like biological specimens, skin cancer, computer tomographic images, and beat-to-beat sequences from electrocardiograms \cite{GVK341772518}. Investigations concerning the Gaussianity of the CMB by applying the SIM to simulated CMB maps, the WMAP 3-year, 5-year or 7-year data were performed in \cite{2003MNRAS.344..115R}, \cite{2007MNRAS.380..466R}, \cite{2009PhRvL.102m1301R, 2009MNRAS.399.1921R} and \cite{2010arXiv1012.2985R}, respectively.

This \textit{Review} is structured as follows: In section \ref{ScalingIndexMethode}, we present the scaling index technique for investigations on a sphere. The data sets that were used for investigations by means of scaling indices to date, including the technique of creating surrogate maps, are introduced in section \ref{DataSets}. The results of these analyses, in particular the detections of non-Gaussianities, asymmetries and local features, are outlined in section \ref{Resultschapter}, followed by the conclusions in section \ref{Conclusions}.


\section{The Scaling Index Method}\label{ScalingIndexMethode}

\begin{figure*}
\centering
\includegraphics[width=17.4cm, keepaspectratio=true, ]{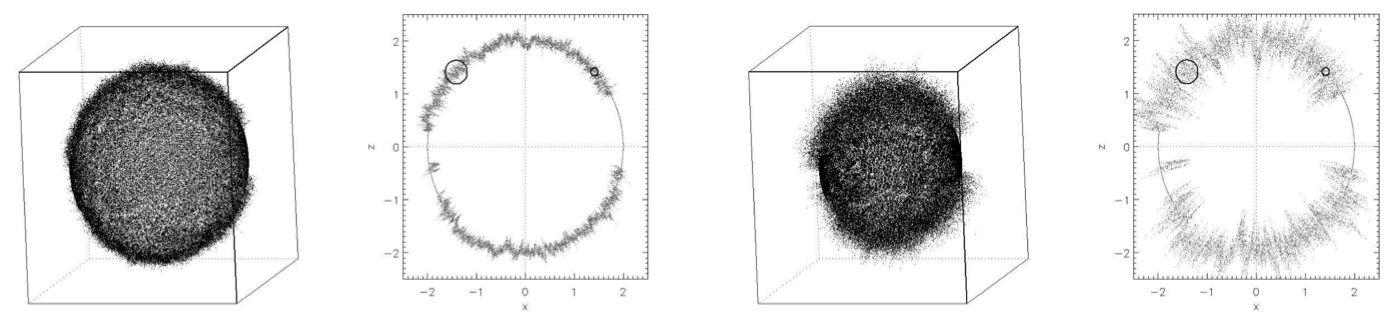}
\caption{WMAP 3-year data after application of the transformation into a three-dimensional point distribution. In the first and the third plot, the full set of points is presented, while the second and fourth shows an $x,z$-projection of only the points with $|y|<0.05$. Two different adjustment parameters were used, $a=0.075$ (on the left) and $a=0.225$ (on the right). The black circles represent the scaling ranges $r=0.075$ and $r=0.225$. Figure taken from \citep{2003MNRAS.344..115R}.} \label{Jitteri}
\end{figure*}

Quite similar to wavelets, weighted scaling indices can be used to perform a \textit{local} analysis of the data set, and can be calculated for different scales, which yields information about characteristic sizes of detected features. The measure has the ability of revealing the topological properties of an input map by detecting different structures in the data, as for example cluster-like or sheet-like structures, as well as filaments or walls. While wavelets are more sensitive to structures, which offer intensity variations of significant magnitude with respect to the existing noise, scaling indices also detect structural features which possess variations within the noise level, but not significantly higher or lower intensity values \cite{2007MNRAS.380..466R}. The SIM turned out to be very useful in the detection of non-Gaussianities in CMB maps \cite{2003MNRAS.344..115R, 2007MNRAS.380..466R, 2009PhRvL.102m1301R, 2009MNRAS.399.1921R, 2010arXiv1012.2985R}.

Scaling indices investigate the spatial distribution of a previously prepared d-dimensional data set. In CMB investigations however, the fluctuations of the temperature maps are characterised by the values of the pixelised sky on a sphere. To be able to apply an analysis by means of scaling indices, one has to combine the temperature information with the two-dimensional spatial information of the map to create a three-dimensional point set, which includes all the information of the original map as spatial information only. This can be done by performing a preprocessing step, namely a transformation of the pixelised spherical sky $S$ into three-dimensional space. Hereby, the pixels $(\theta_i,\phi_i)$, $i=1,...,N_{pix}$, of $S$, where $N_{pix}$ denotes the number of pixels and $(\theta_i,\phi_i)$ latitude and longitude of the pixel $i$ on the sphere, are converted into a three-dimensional \textit{jitter}: Each temperature value $T(\theta_i,\phi_i)$ is assigned to one point $\vec{p}_i$, which is located in the radial direction through its pixel´s centre $(\theta_i,\phi_i)$, that is a straight line perpendicular to the surface of the sphere. Thus, the three-dimensional position vector of the new point $\vec{p}_i$ reads as
\begin{align*}
x_i &= (R+dR) \cos(\phi_i) \sin(\theta_i) \\
y_i &= (R+dR) \sin(\phi_i) \sin(\theta_i) \\
z_i &= (R+dR) \sin(\theta_i)
\end{align*}
with
\[ dR = a \left( \frac{T(\theta_i,\phi_i) - \langle T \rangle}{\sigma_T} \right) \]
with $R$ denoting the radius of the sphere and $a$ describing an adjustment parameter. In addition, $\langle T \rangle$ and $\sigma_T$ characterise the mean and the standard deviation of the temperature fluctuations, respectively. The normalisation is performed to obtain for $dR$ zero mean and a standard deviation of $a$. It is recommended to choose both $R$ and $a$ in a proper way to ensure a high sensitivity of the SIM with respect to the temperature fluctuations at a certain spatial scale. For CMB analysis, it turned out that this requirement is provided using $R=2$ for the radius of the sphere and setting the adjustment parameter $a$ to the value of the below introduced scaling range parameter $r$ \cite{2007MNRAS.380..466R}. A CMB map transformed to a three-dimensional point distribution is illustrated in Figure \ref{Jitteri}. Hereby, two different adjustment parameters $a$ were used in the embedding process.

\begin{figure*}
\centering
\includegraphics[width=17.4cm, keepaspectratio=true, ]{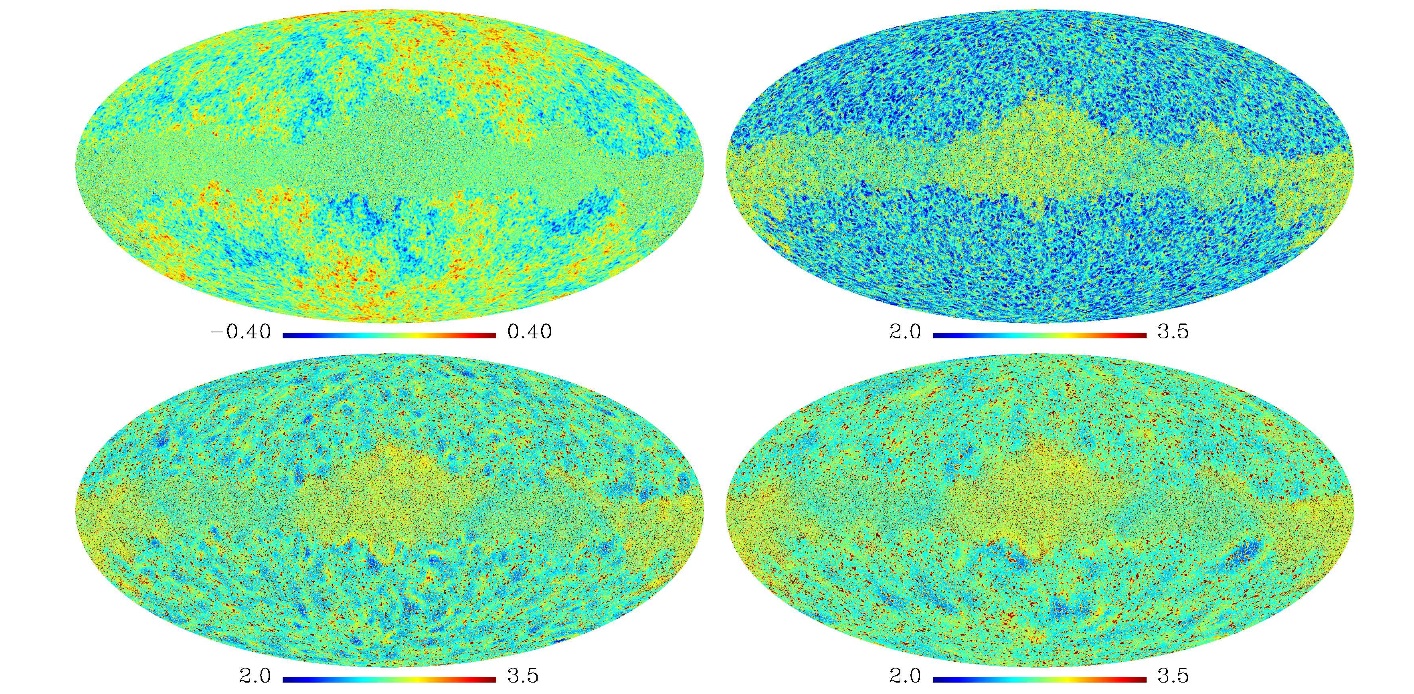}
\caption{A simulated CMB map, in which the central regions were masked out and filled with nearly white noise, whereby the spatial noise patterns are preserved (see section \ref{WMAPi}) (upper left), and its scaling index responses $\alpha(\vec{p}_i,r)$ for three different scaling ranges: $r=0.05$ (upper right), $r=0.15$ (lower left) and $r=0.25$ (lower right). Different values of $\alpha(\vec{p}_i,r)$ correspond to different types of structure in the underlying map. Small scaling ranges examine the behaviour of the small structures, while the characteristics of the larger structure is displayed by the higher scaling ranges. Note the different structure inside and outside the masked region of the simulated map, which is clearly identified by the scaling indices. These maps (and all following ones) are shown in a conventional scheme, namely the Mollweide projection in the Galactic reference frame with the Galactic Centre at the centre of the image and the longitude increasing from there to the left-hand side.} \label{Alphi}
\end{figure*}

After this preprocessing step, the actual scaling index technique can be applied. In general, the SIM is a mapping that calculates for every point $\vec{p}_i$ of the point set $P$ a single value, which depends on the spatial position of $\vec{p}_i$ in the group of the other points in which $\vec{p}_i$ is embedded in. As already stated above, $P$ is three-dimensional for the case of CMB analysis. For every point $\vec{p}_i$, the \textit{local weighted cumulative point distribution} is defined as
\[ \rho(\vec{p}_i,r) = \sum_{j=1}^{N_{pix}} s_r(d(\vec{p}_i,\vec{p}_j)) \]
with $r$ describing the scaling range, while $s_r(\bullet)$ and $d(\bullet)$ denote a shaping function and a distance measure, respectively. To obtain our measure, the \textit{scaling index} $\alpha(\vec{p}_i,r)$, we assume the following scaling law:
\[ \rho(\vec{p}_i,r) \propto r^{\alpha(\vec{p}_i,r)} . \]
The scaling index can therefore be computed as the logarithmic derivative of $\rho(\vec{p}_i,r)$. Formally, this reads as
\[ \alpha(\vec{p}_i,r) = \frac{\partial \log \rho(\vec{p}_i,r)}{\partial \log r} . \]
In general, one is free to choose the shape of $s_r(\bullet)$ and $d(\bullet)$. For the recent analyses that are discussed in this review, a set of quadratic gaussian shaping functions as well as the Euclidian norm were applied:
\begin{align*}
s_r(x) &= e^{-(\frac{x}{r})^2} \ , \\
d(\vec{p}_i,\vec{p}_j) &= \| \vec{p}_i - \vec{p}_j \| \ .
\end{align*}
Taking this into account, and using in addition the abbreviation $d_{ij} := d(\vec{p}_i,\vec{p}_j)$, we obtain the final formula of the scaling indices:
\begin{equation}\label{AlphaFormel1}
\alpha(\vec{p}_i,r) = \frac{\sum_{j=1}^{N_{pix}} 2\big(\frac{d_{ij}}{r}\big) e^{-\big(\frac{d_{ij}}{r}\big)^2}}{\sum_{j=1}^{N_{pix}} e^{-\big(\frac{d_{ij}}{r}\big)^2}}.
\end{equation}
In the resulting map $\alpha(\vec{p}_i,r)$, $i=1,...,N_{pix}$, the structural behaviour of the underlying point set $P$ becomes apparent, and different types of structure can be detected very easily. The values of $\alpha$ are related to structural characteristics in the following way: A point- or cluster-like structure leads to scaling indices $\alpha \approx 0$, filaments to $\alpha \approx 1$ and sheet-like structures to $\alpha \approx 2$. A uniform distribution of points would result in $\alpha \approx 3$. In between, curvy lines and curvy sheets produce $1 \leq \alpha \leq 2$ and $2 \leq \alpha \leq 3$, respectively. Underdense regions in the vicinity of point-like structures, filaments or walls feature $\alpha > 3$. An example of a simulated CMB map and its scaling index response is shown in Figure \ref{Alphi}.

From equation (\ref{AlphaFormel1}), one can see that the scaling range parameter $r$ can be chosen arbitrarily. This parameter weights the distances between our point of interest $\vec{p}_i$ and the remaining points $\vec{p}_j$ (see also definition of $s_r(x)$). Therefore, we can make use of smaller or larger values for $r$ to examine the different behaviour of the small-scale or large-scale structural configuration in the underlying map. For the analyses that were done by means of the scaling indices so far (see section \ref{Resultschapter}), it was common to make use of the ten scaling range parameters $r_k = 0.025, 0.05, ..., 0.25$, $k=1,2,...10$. Figure \ref{Alphi} gives an example for the results of the SIM for three different values of $r$, applied onto a simulated CMB map.


\section{Data Sets}\label{DataSets}

In the publications concerning scaling index analysis with CMB to date, different WMAP data sets as well as different techniques to handle foreground contaminated regions were applied. To test for the amount of non-Gaussianity in the WMAP data, simulations based on Gaussian random fields were constructed, which is the most common way in CMB analysis. In addition, so-called surrogate maps were generated. With the help of these surrogates, it is possible to test for the more specific hypothesis of uncorrelated phases. The surrogate method also offers the possibility to analyse the data in a scale-dependent but model-\textit{in}dependent way. In the following, we will give an overview of the used sets of maps.

\begin{figure*}
\centering
\includegraphics[width=17.4cm, keepaspectratio=true, ]{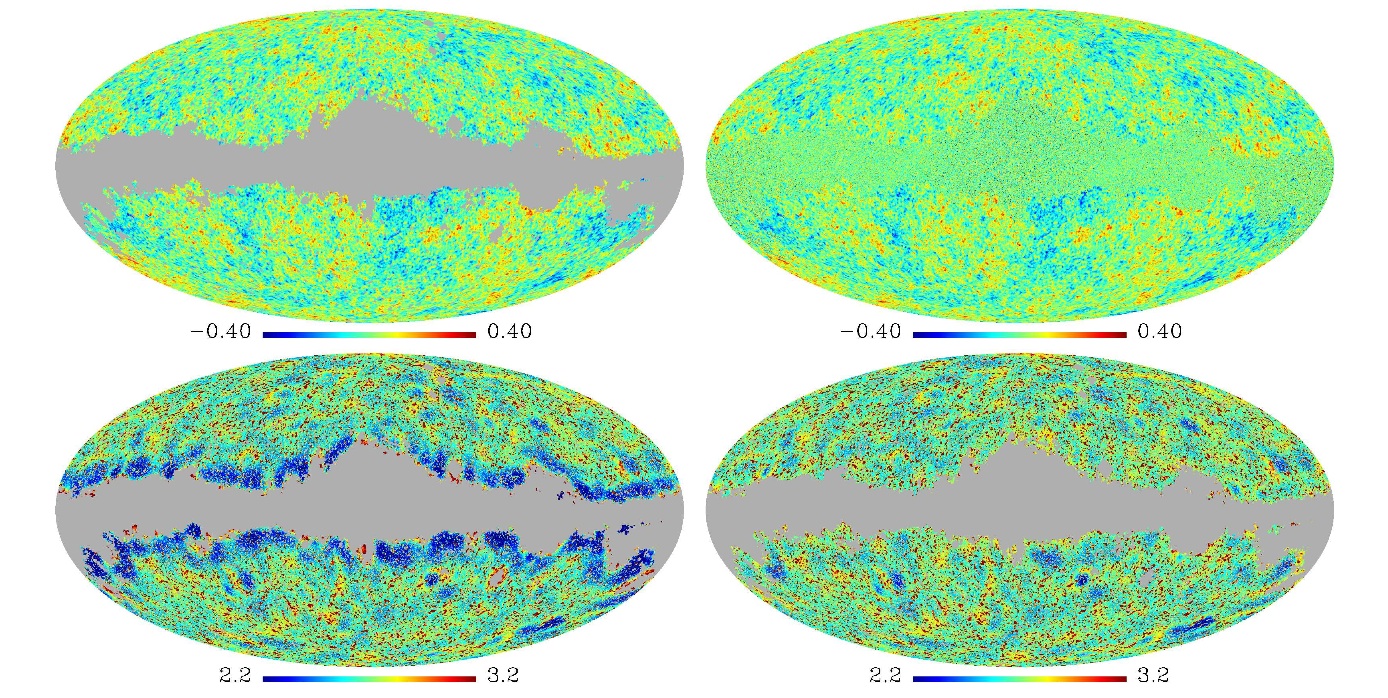}
\caption{The two plots on the left hand side illustrate the original 5-year WMAP data of the co-added VW-band (above) and the related colour-coded $\alpha$-response (below) for a scaling range of $r = 0.175$. The Galactic Plane and several secondary spots are masked out by the KQ75-mask \cite{2011ApJS..192...15G}. The equivalent plots for the mask-filling technique are arranged on the right hand side.} \label{fig1:dreimaldrei}
\end{figure*}

\subsection{The WMAP data}\label{WMAPi}

In the current fourth data release of the WMAP team\footnote{http://lambda.gsfc.nasa.gov} \cite{2011ApJS..192...14J}, the recording of seven years of observation were made publicly available. One of the main challenges in all data releases as well as subsequent investigations is the handling of the heavily foreground contaminated regions, caused by different point sources and in particular the Galactic plane \cite{2011ApJS..192...15G}. These approaches result in different maps, whereof the following ones were subject to an analysis by means of scaling indices:\par
1.) \textit{The ILC map}\par
The 7-years foreground-cleaned internal linear combination (ILC) map \cite{2011ApJS..192...15G} is generated and provided by the WMAP team (in the following: ILC7). Measurements of all observed frequency bands are combined in terms of a linear combination. Different weights for the different bands as well as for different previously chosen fractions on the sky are determined to minimise the variance of the temperature fluctuations. The ILC7 map is downgraded to a resolution of 786432 pixels, which equals to $N_{side} =  256$ in the employed HEALPix-software\footnote{http://healpix.jpl.nasa.gov} \cite{2005ApJ...622..759G}.\\
2.) \textit{The Needlet-based ILC map}\par
For comparison we also included the map produced in \cite{2009A&A...493..835D}, namely the 5-years needlet-based ILC map, which has been shown to be significantly less contaminated by foreground and noise than other existing maps obtained from WMAP data (in the following: NILC5). The NILC5 map is downgraded to $N_{side} =  256$ as well.\par
3.) \textit{The masked band-wise maps}\par
Unlike to the two ILC maps from above, the single Q-, V- and W-bands of the WMAP-satellite as well as a co-added VW-map can shed light on the influence of the different wavelenght depending foregrounds onto the CMB signal. Although we work with the maps that are reduced by means of the Foreground Template Model proposed in \cite{2007ApJS..170..288H} and \cite{2007ApJS..170..335P}, these maps still show strong foreground effects and differ from each other, making a band-wise analysis reasonable. To obtain the band-wise or combined maps, we accumulate the differencing assemblies Q1, Q2, V1, V2, W1, W2, W3, W4 via a noise-weighted sum \citep{2003ApJS..148....1B}:
\begin{equation} \label{GewSum}
T(\theta, \phi) = \frac{\sum_{i \in \mathcal{A}} T_i(\theta,\phi)/ \sigma^2_{0,i} } {\sum_{i \in \mathcal{A}} 1/ \sigma^2_{0,i}}
\end{equation}
In this equation, $\mathcal{A}$ characterises the set of required assemblies, e.g. for the co-added VW-map $\mathcal{A} = \lbrace V1,V2,W1,W2,W3,W4 \rbrace$, while the noise per observation of the different assemblies, given in \cite{2009ApJS..180..225H}, is denoted by $\sigma_0$. The co-added VW-map is created for the 3-year and the 5-year data, while the single Q-, V- and W-band maps are generated for the 5-year data only. \par
As for the ILC maps above, we decrease the resolution to $N_{side} =  256$. In addition, the heavily foreground-affected parts of the sky are cut out, using the Kp0-mask for the WMAP 3-year, and the KQ75-mask for the WMAP 5-year observations \citep{2009ApJS..180..265G}. Hereby, both of them have to be downgraded as well. We choose a conservative downgrading of the mask by taking only all pixels at $N_{side} =  256$ that consist completely of non-mask-pixels at $N_{side} =  512$. All downgraded pixels at $N_{side} =  256$, for which one or more pixels at $N_{side} =  512$ belonged to the Kp0- or KQ75-mask, respectively, are considered to be part of the downgraded mask as well. In doing so, $23,5\%$ of the sky is removed for the Kp0-mask, while the KQ75-mask is even more conservative with $28.4\%$ (see upper left part of Figure \ref{fig1:dreimaldrei}). Finally, we remove the residual monopole and dipole by means of the appropriate HEALPix routine applied to the unmasked pixels only.\par
4.) \textit{The mask-filled band-wise maps}\par
\label{ch:maskfill}
Just cutting out the masked regions like above spoils the results of the scaling index method: Instead of a more or less uniform distribution, the $\alpha$-values in the regions around the mask now detect a sharp boundary with no points in the masked area, into which the scaling regions extend (see Figure \ref{Jitteri}). This results in lower values of $\alpha$. The effect can clearly be seen in the $\alpha$-response of the masked VW-band 5-year WMAP-data in the lower left corner of Figure \ref{fig1:dreimaldrei}. A solution to this problem is to \textit{fill} the masked areas with suitable values, that prevent the low outcome at the edges of the mask. This can be accomplished by applying the following two steps: \par
At first, we fill the masked regions with Gaussian noise, whose standard deviation for each pixel corresponds to the pixel noise made available by the WMAP-team:
\[ T_{mask}^{\ast}(\theta, \phi) \sim \mathcal{N}( 0 , \sigma_{(\theta, \phi)}^2 ) \]
Here, $\sigma_{(\theta, \phi)}$ denotes the pixel noise of the pixel which is located in the direction $(\theta, \phi)$. Then, we scale the expectation value and the variance as a whole to the empirical mean $\mu_{rem}$ and variance $\sigma_{rem}^2$ of the remaining regions of the original temperature map:
\[ T_{mask}(\theta, \phi) = \frac{\sigma_{rem}^2}{\sigma_{mask}^2} \ T_{mask}^{\ast}(\theta, \phi) + \mu_{rem} \]
with
\begin{align*} \label{Fuellformel1_zusatz}
\mu_{rem} \ \ &= \ \frac{1}{N_\mathcal{R}} \sum_{(\theta, \phi) \in \mathcal{R}} T(\theta, \phi)  \\
\sigma_{rem}^2 \ \ &= \ \frac{1}{N_\mathcal{R}-1}\sum_{(\theta, \phi) \in \mathcal{R}} (T(\theta, \phi)-  \mu_{rem})^2  \\
\sigma_{mask}^2 \ &= \ \frac{1}{N_\mathcal{M}-1}\sum_{(\theta, \phi) \in \mathcal{M}} T_{mask}^{\ast}(\theta, \phi)^2
\end{align*}
where $\mathcal{R}$ and $\mathcal{M}$ stand for the non-masked and masked region of the map respectively, and $N_\mathcal{R}$ as well as $N_\mathcal{M}$ denote their number of pixels. Thus, we filled the mask with (nearly) white Gaussian noise whose mean and standard deviation equal the respective terms of the remaining map, whereby the spatial noise patterns are preserved. With this filling technique, we obtain a complemented data set instead of just excluding the masked regions. Boundary effects caused by the mask can be eliminated. The right column of Figure \ref{fig1:dreimaldrei} shows the filling method as well as the corresponding $\alpha$-response.

\subsection{Simulations}

A simple approach to evaluate the amount of non-Gaussianity in the WMAP data is to compare the measured data with maps that fulfil the Gaussian hypothesis. For the band-wise analysis, it is important to create simulations for each respective band. The proceeding is hereby as follows: We take the best fit $\Lambda CDM$ power spectrum $C_l$, derived from the respective WMAP 3-year or 5-year data only, and the according window function for each differencing assembly (Q1-Q2, V1-V2, W1-W4), as again made available on the LAMBDA-website. With these requisites, one can create Gaussian random fields mimicking the Gaussian properties of the best fit $\Lambda CDM$-model and including the WMAP-specific beam properties by convolving the $C_l$´s with the window function. For every assembly, we add Gaussian noise to these maps with a particular variance for every pixel of the sphere. This variance depends on the number of observations $N_i(\theta,\phi)$ in the respective direction and the noise dispersion per observation, $\sigma_{0,i}$. After this procedure, the co-added VW-band (for the 3-year and 5-year analyses) as well as the Q-, V- and W-bands (for the 5-year investigations only) can be summarised using equation (\ref{GewSum}) from above and decreased to the resolution of $N_{side} =  256$. The respective Kp0- or KQ75-mask is cut out and the residual monopole and dipole removed, just as for the WMAP-data above. For comparison with the mask-filled data maps, the filling method described above is additionally applied onto the 5-year simulations as well.

\subsection{Surrogates}

\begin{figure*}
\centering
\includegraphics[width=8cm, keepaspectratio=true, ]{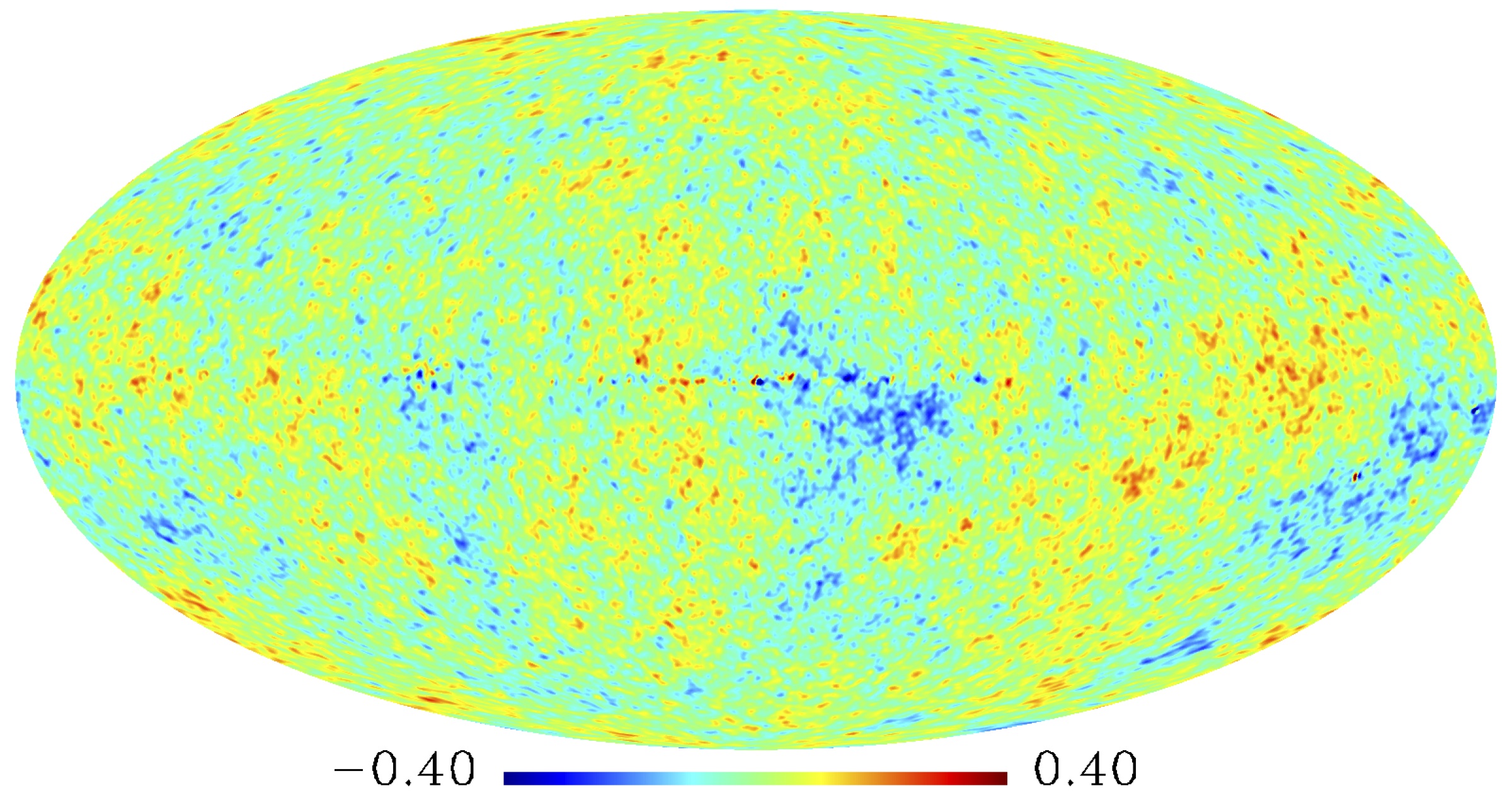}
\includegraphics[width=8cm, keepaspectratio=true, ]{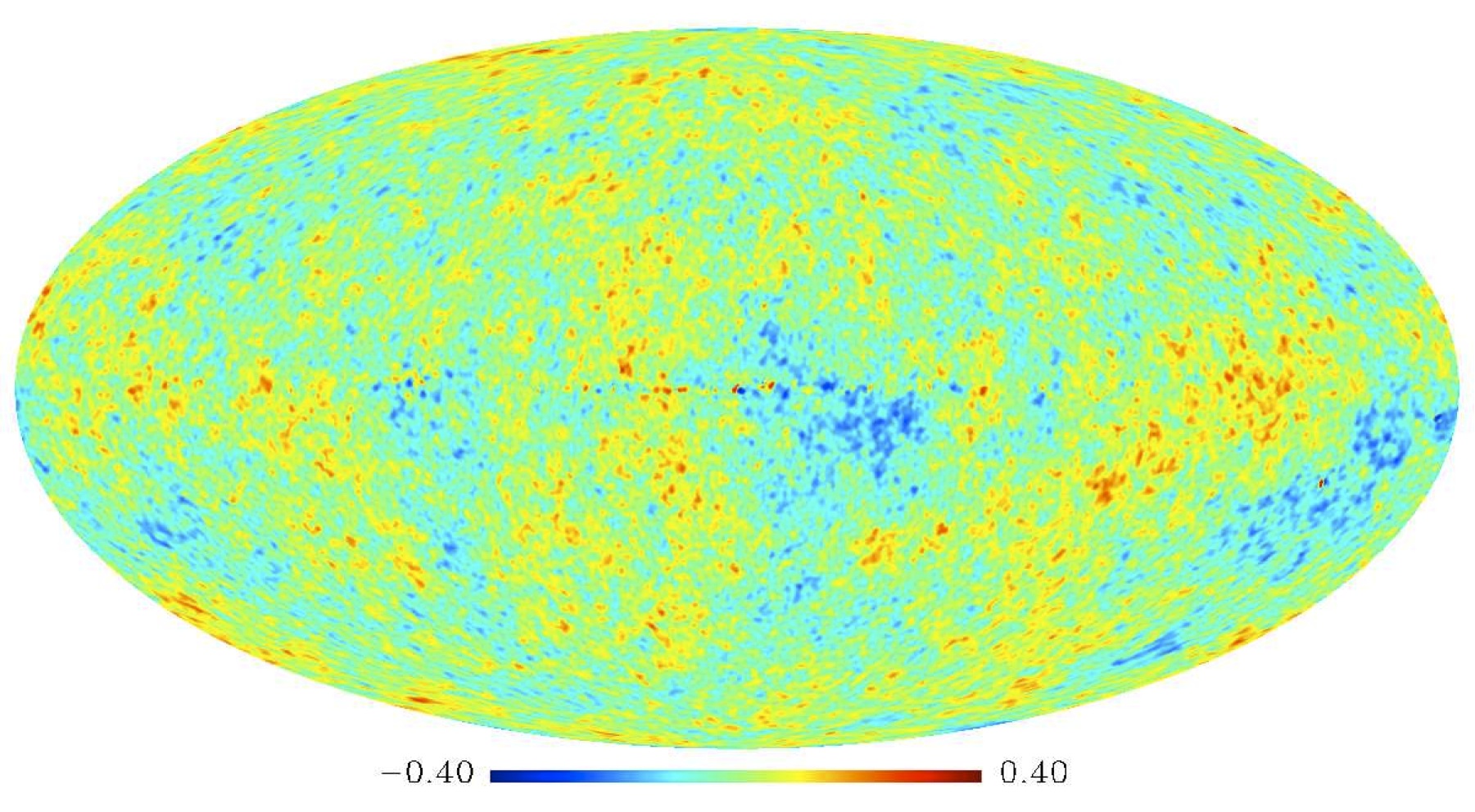}
\includegraphics[width=8cm, keepaspectratio=true, ]{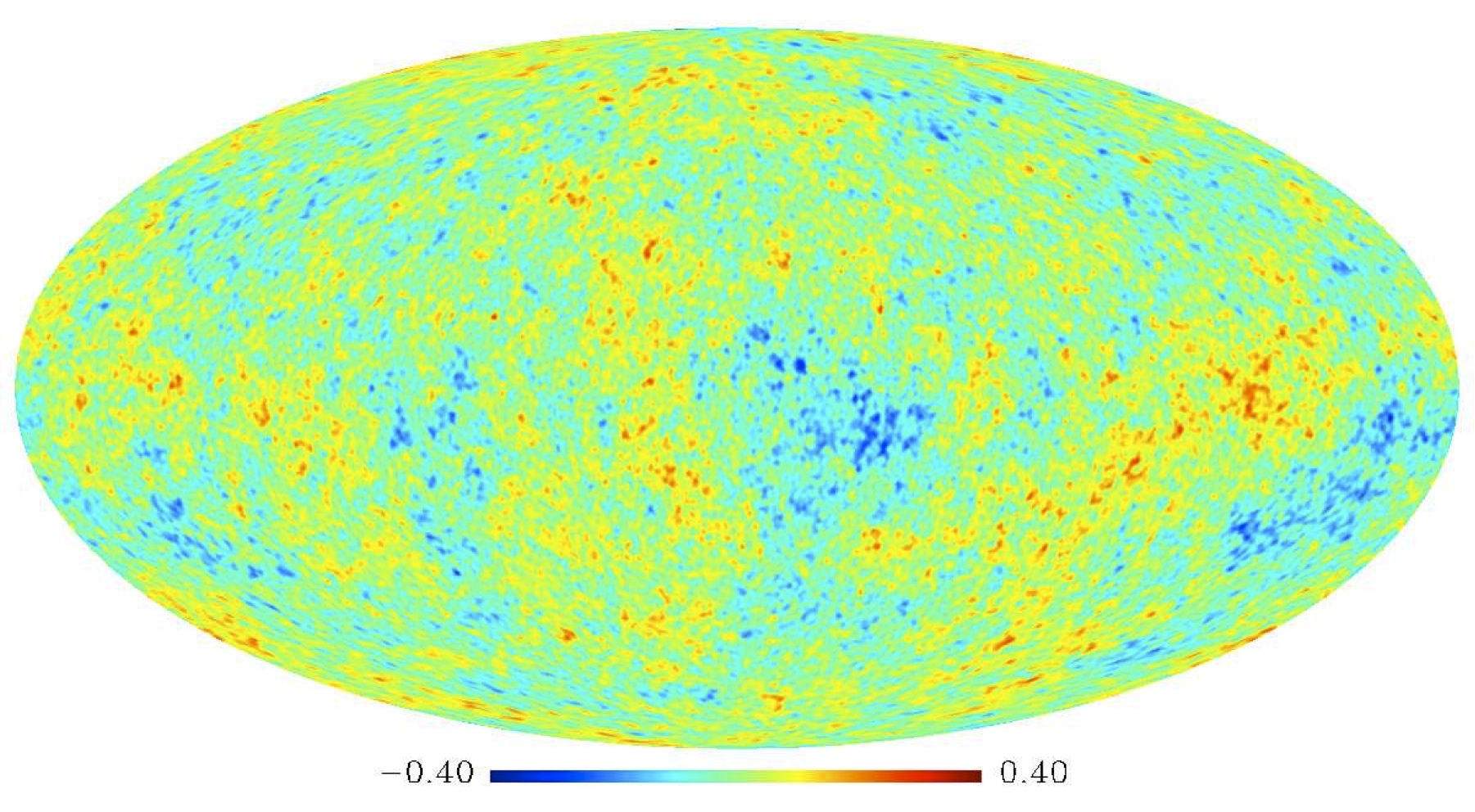}
\includegraphics[width=8cm, keepaspectratio=true, ]{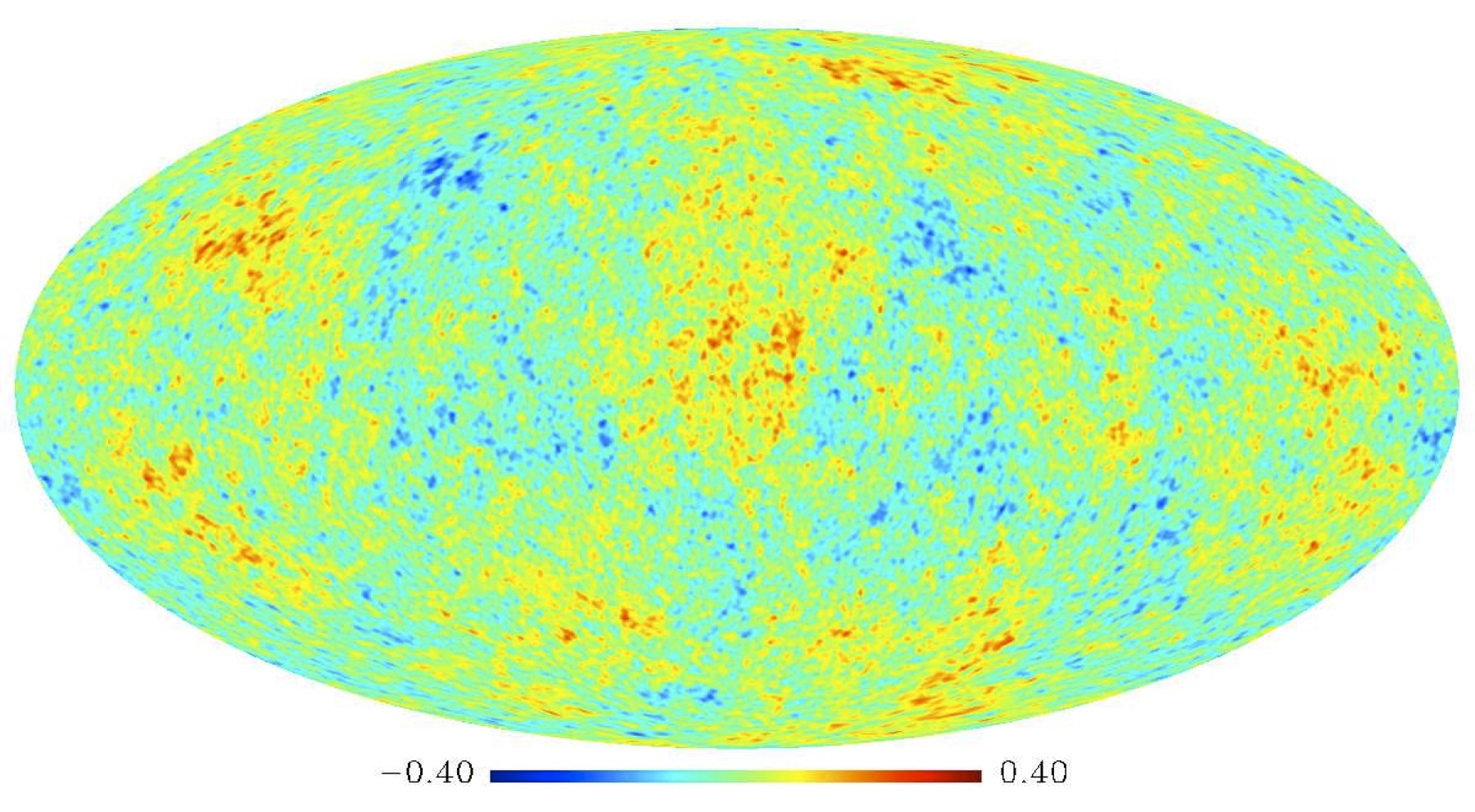}
\caption{The ILC map in its original form (upper left) and after remapping of the temperatures and phases (upper right). First order (lower left) and respective second order surrogate (lower right) for $l_{cut}=20$. Note the resemblance of the remapped with the ILC map for all, but also of the first order surrogate with the ILC map at large scales.} \label{Surrogateplot}
\end{figure*}

A comparison with simulated CMB maps represents the most obvious and common approach to search for non-Gaussianities in the data set. However, it is also possible to create maps, so-called surrogates, that are similar to the original map except for one (or more) previously selected feature(s) which is (are) randomised. By comparing the data with this set of maps, one focuses on the deviations caused by the randomisation of these feature(s). The whole proceeding is therefore \textit{model-independent}. One way of applying this method in terms of a scale-dependent search for non-Gaussianities has been proposed and discussed in \cite{2009PhRvL.102m1301R} and \cite{2010arXiv1012.2985R}. In the following, we describe the various steps for generating surrogate maps in more detail:

Consider a CMB map $T(\theta,\phi)$, where $T(\theta,\phi)$ is Gaussian distributed, and its Fourier transform. The complex valued Fourier coefficients $a_{\ell m}$ can be written as
$a_{\ell m} = | a_{\ell m} | e^{i \phi_{\ell m}} $ 
with  $\phi_{\ell m}=\arctan \left( {\rm Im}(a_{\ell m}) / {\rm Re}(a_{\ell m} )  \right)$. The linear or Gaussian properties of the underlying random field are contained in the absolute values $ | a_{\ell m} | $, whereas all higher-order correlations (HOCs) -- if present -- are encoded in the phases $\phi_{\ell m}$ and the correlations among them. Having this in mind, a versatile approach for testing for scale dependent non-Gaussianities relies on a scale-dependent shuffling procedure of the phase correlations followed by a statistical comparison of the so-generated surrogate maps.

However, the Gaussian shape of the histogram of the temperature distribution and the randomness of the set of Fourier phases in the sense that they are uniformly distributed in the interval $[- \pi, \pi]$, are a necessary prerequisite for the application of the surrogate-generating algorithm, which we propose in the following. To fulfil these two conditions, we perform the following preprocessing steps. First, the maps are remapped onto a Gaussian distribution in a rank-ordered way. This means that the amplitude distribution of the original temperature map in real space is replaced by a Gaussian distribution in a way that the rank-ordering is preserved, i.e. the lowest value of the original distribution is replaced with the lowest value of the Gaussian distribution etc. By applying this remapping we automatically focus on HOCs induced by the spatial correlations in the data while excluding any effects coming from deviations of the temperature distribution from a Gaussian one.

To ensure the randomness of the set of Fourier phases we performed a rank-ordered remapping of the phases onto a set of uniformly distributed ones followed by an inverse Fourier transformation. These two preprocessing steps only have marginal influence to the maps (see Figure \ref{Surrogateplot}). The main effect is that the outliers in the temperature distribution are removed. Due to the large number of temperature values (and phases) we did not find any significant dependence of the specific Gaussian (uniform) realisation used for remapping of the temperatures (phases). The resulting maps may already be considered as a surrogate map and we named it zeroth order surrogate map. The first and second order surrogate maps are obtained as follows:

At first, one generates a first order surrogate map, in which any phase correlations for the scales, which are not of interest, are randomised. This is achieved by a random shuffle of the phases $\phi_{\ell m}$ for $\ell \notin  \Delta \ell = [\ell_{min}, \ell_{max}], \ 0 < m \le \ell$ and by performing an inverse Fourier transformation.

In a second step, a chosen number of realisations of second order surrogate maps are generated for the first order surrogate map, in which the remaining phases $\phi_{\ell m}$  with $\ell \in \Delta \ell$, $0 < m \le \ell$ are shuffled, while the already randomised phases for the scales, which are not under consideration, are preserved. Note that the Gaussian properties of the maps, which are given by $| a_{\ell m} |$, are {\it exactly} preserved in all surrogate maps.

In \cite{2009PhRvL.102m1301R}, the surrogates method was applied only to the $\ell$-range  $\Delta \ell = [2, 20]$, while in \cite{2010arXiv1012.2985R}, the analysis were extended to smaller scales as well: Three more $\ell$-intervals, namely $\Delta \ell = [20, 60]$, $\Delta \ell = [60, 120]$ and $\Delta \ell = [120, 300]$ were considered. The choice of $60$ as $\ell_{min}$ and  $\ell_{max}$ is somewhat arbitrary, whereas the $\ell_{min}=120$ and  $\ell_{max}=300$ for the last $\ell$-interval was selected in such a way that the first peak in the power spectrum is covered. Going to even higher $\ell$ does not make much sense, because the ILC7 map is smoothed to one degree FWHM. In principle, one could include higher $\ell$, since some other maps -- especially NILC5 -- are not smoothed. But to allow for a consistent comparison of the results obtained with the different observed and simulated input maps, a restriction to only investigate $\ell$-intervals up to  $\ell_{max}=300$ is applied.

Besides this two-step procedure aiming at a dedicated scale-dependent search for non-Gaussianity, one can also test for non-Gaussianity using surrogate maps without specifying certain scales. In this case there are no scales, which are not of interest, and the first step in the surrogate map making procedure becomes dispensable. The zeroth order surrogate map is to be considered here as first order surrogate and the second order surrogates are generated by shuffling all phases with $0 < m \le \ell$ for all available $\ell$, i.e. in our case $\Delta \ell = [2, 1024]$.

Finally, for calculating scaling indices to test for higher order correlations, the surrogate maps were degraded to $N_{side}=256$ and residual monopole and dipole contributions were subtracted. In contrast to a comparison between the data and simulated maps, which reveals all kinds of deviations from Gaussian random fields, the statistical comparison of the two classes of surrogates focusses on \textit{possible HOCs on certain scales}, and the question if these have left traces in the first order surrogate maps, which were then deleted in the second order surrogates.


\section{Results}\label{Resultschapter}

\subsection{Comparison with Simulations}

\subsubsection{The WMAP three-year data}
\label{ch:wmap3}
\begin{figure}
\centering
\includegraphics[width=8cm, keepaspectratio=true, ]{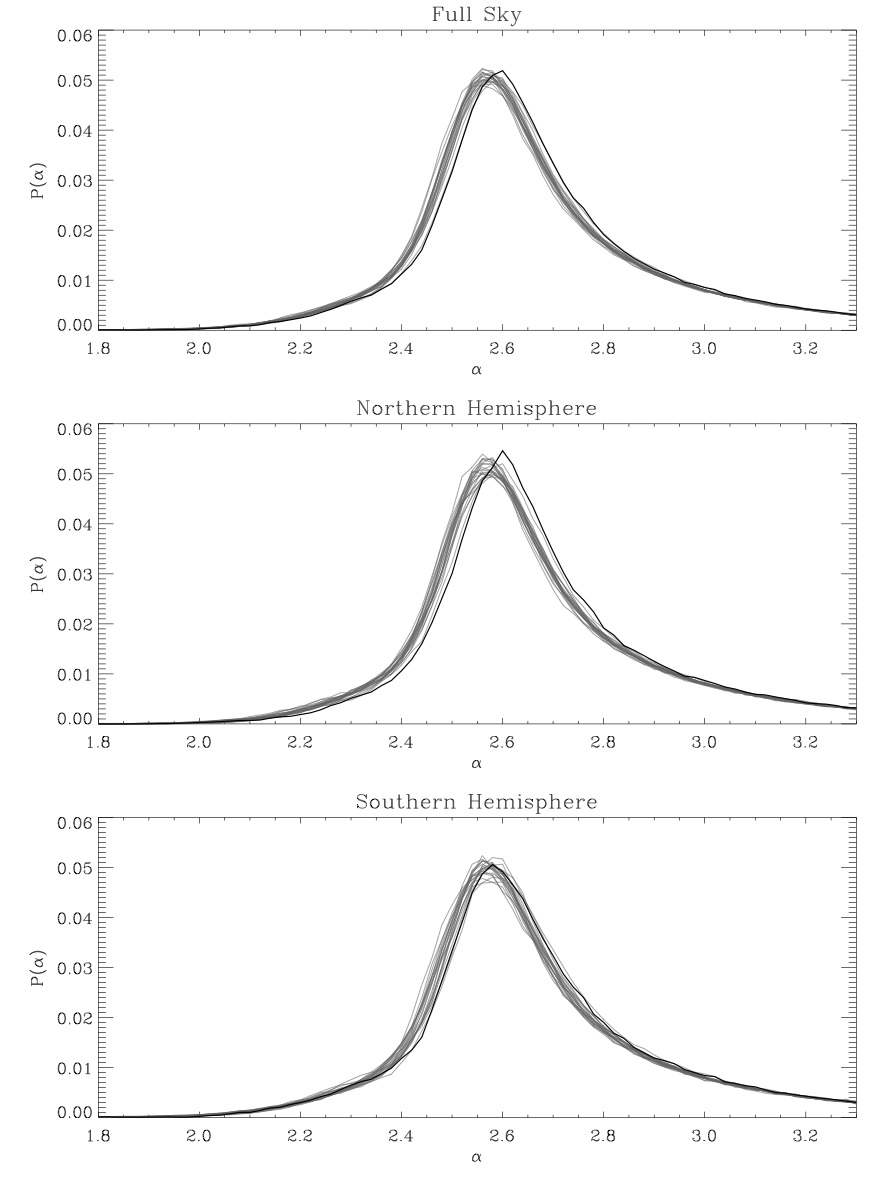}
\caption{Probability density $P(\alpha)$ for the co-added WMAP 3-year data (black) and 20 simulations (grey) for $r=0.175$ and the Kp0-mask.}
\label{fig5:probWMAP3}
\end{figure}

Finding differences between observed and simulated CMB maps which fulfil the Gaussian hypothesis of the best fitting $\Lambda$CDM model is a strong indication of the existence of non-Gaussianities in the CMB. The WMAP 3-year data was compared to $N=1000$ simulations. The probability densities $P(\alpha)$ of the scaling indices for one selected scale ($r=0.175$) are displayed in Figure \ref{fig5:probWMAP3} for the WMAP 3-year data and a subset of 20 simulations. 
The probability density for the WMAP data is shifted towards higher values, which indicates that the underlying temperature fluctuations for the observed data resemble more 'unstructured', that is, random and uniform fluctuations in comparison to the simulations. This effect is more pronounced in the northern hemisphere of the galactic coordinate system than in the southern. Furthermore, the histograms of the simulations are slightly broader, indicating that the simulations exhibit a larger structural variability than the observed data. 

\begin{figure}
\centering
\includegraphics[scale =0.5,angle=0]{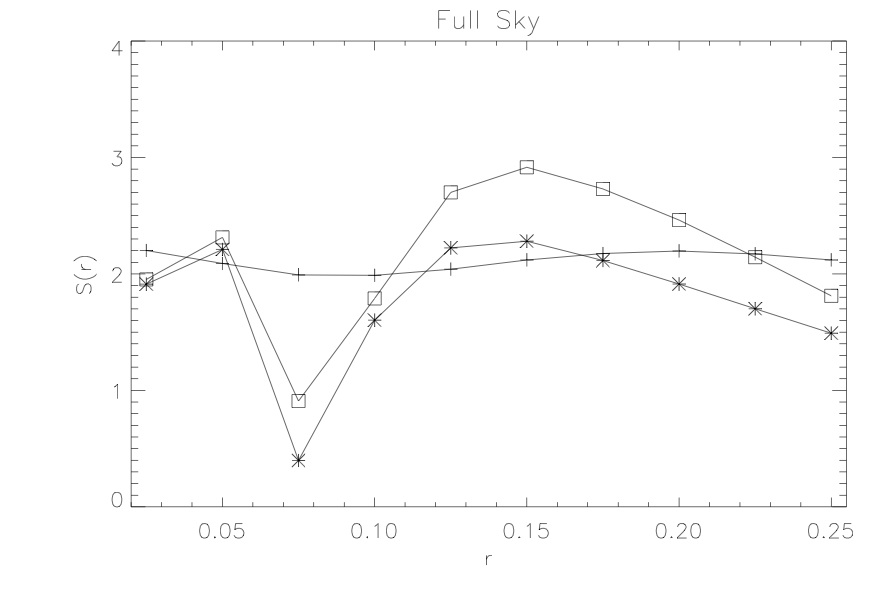}
\includegraphics[scale =0.5,angle=0]{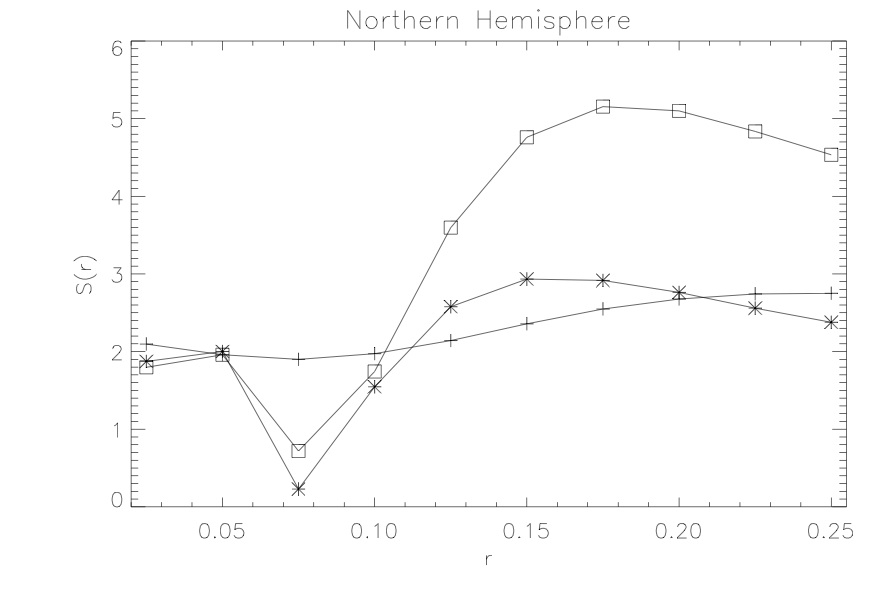}
\includegraphics[scale =0.5,angle=0]{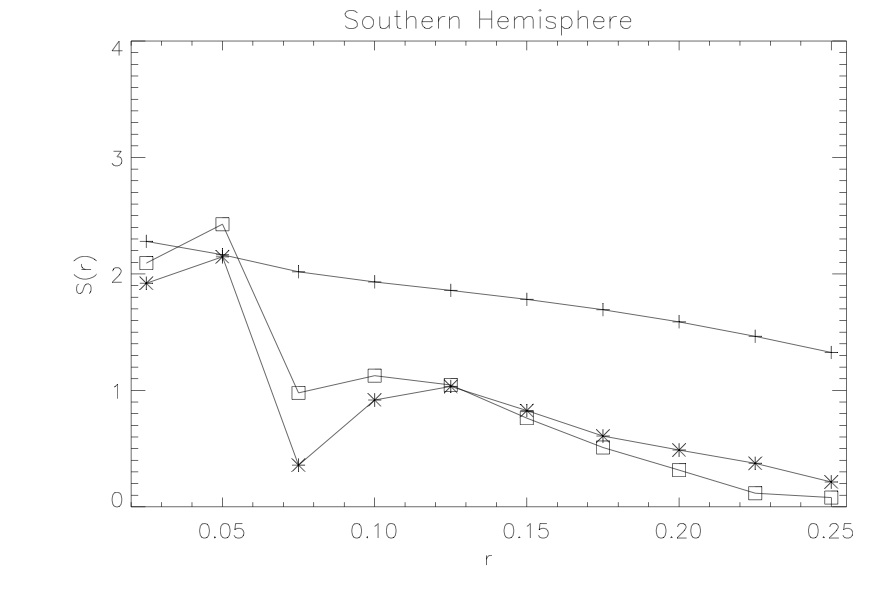}
\caption{Deviations of the (combined) moments of the $\alpha$ distribution of the WMAP 3-year data with the Kp0-mask as a function of the scaling range $r$. The ‘+’ symbol 
denotes the mean, the ‘$\ast$’ symbol denotes the standard deviation and the boxes denote the $\chi^2$ combination of the mean and standard deviations.}
\label{fig6:devWMAP3}
\end{figure}
These effects can more exactly be quantified by calculating the mean and  
standard deviation for the distribution of scaling indices as calculated for different 
scaling ranges. For scales larger than $r=0.1$ the mean of the 
scaling indices is always systematically higher for WMAP 
than for the simulations. The effect is much more pronounced in the northern hemisphere. 
For the standard deviation we observe for the same scales significantly lower values for WMAP 
in the northern hemisphere  and slightly higher ones for the southern sky. 
For the full sky these two effects cancel each other so that the observed deviations 
to lower values are no longer so significant.\\
Besides the mean and standard deviation we additionally considered a combination 
of these two test statistics, namely a diagonal $\chi^2$-statistic at a given scale $r_k$,

\begin{equation} \label{Chi2}
\chi^2_{\langle \alpha(r_k) \rangle,\sigma_{\alpha(r_k)}} = \sum \limits_{j=1}^{2} \left[ \frac{M_j - \langle M_j\rangle }{\sigma_{M_j}} \right]^2,
\end{equation}

where $M_1(r_k) = \langle \alpha(r_k) \rangle$, $M_2(r_k) = \sigma_{\alpha(r_k)}$. These statistics are computed for both the simulations and the observed moments. The $\sigma$-normalised deviations  $S$ of the WMAP data from the simulations
\begin{equation}\label{eq:significance}
S(Y) = \frac{Y_{data} - \langle Y_{simulation} \rangle}{\sigma_{Y_{simulation}}}
\end{equation}

(with $Y=\langle \alpha(r_k) \rangle$, $\sigma_{\alpha(r_k)}$, $\chi^2$) are shown in Figure  \ref{fig6:devWMAP3}. The mean $\langle$ $\rangle$ and standard deviation $\sigma$ are obtained by summing over $N=1000$ simulations.
The percentages $p$ of the simulations with higher (lower, respectively) results of the scale-independent diagonal $\chi^2$-statistics are calculated as well. The levels for the detection of non-Gaussianities (NGLs) are very high and do not fall below 99\% for any scale. Even higher values for both the significances and the confidence levels are found, if one only considers the northern hemisphere. For scales larger than $r=0.15$ none of the simulations was found to have a higher values for $\langle \alpha \rangle$ than the observation. For the southern hemisphere, however, both the significances and the confidence levels for the smaller radii are slightly higher than for the northern sky but continuously decrease for higher radii $r$. \\
For the standard deviation we find slightly different results. In a transition regime $r \approx 0.075$ the width of the distribution of $\alpha$ is practically the same for the observation and the Monte Carlo sample. On smaller scales $\sigma_{\alpha}$ is higher for WMAP, on larger scales $r$ we observe higher standard deviations for the simulations. This effect is more pronounced in the northern hemisphere. For the largest scales the differences for $\sigma$ between simulations and observation diminishes. Especially, for the southern hemisphere no signatures for deviations from Gaussianity are identified at larger scales using $\sigma_{\alpha}$.
The behaviour of the $\chi^2$- statistics as a function of the scale parameters $r$  can -- as expected -- be regarded as
a superposition of the two underlying statistics $<\alpha>$ and $\sigma_{\alpha}$.

Some readers might argue that the selection of certain moments (mean, standard deviation, $\chi^2$) and scales $r_k$ for highest significance, represents an \textit{a posteriori} choice analysing the data. Although a choice might be well motivated by the results obtained with simulations, we are also using statistics that are \textit{a priori}. In order to test for non-Gaussianity, we calculated scale-independent diagonal $\chi^2$ statistics, where we considered only one (mean or standard deviation) or both measures, and summed over $N_r$ considered length scales $r_k$, $k=1,...,10$.  

\begin{equation}
\chi^2_{\langle \alpha \rangle}= \sum_{k=1}^{N_r}   \left[ \frac{M_1(r_k) - \langle M_1(r_k) \rangle}{\sigma_{M_1(r_k)}} \right]^2
\end{equation}

\begin{equation}
\chi^2_{\sigma_{\alpha}}= \sum_{k=1}^{N_r}   \left[ \frac{M_2(r_k) - \langle M_2(r_k) \rangle}{\sigma_{M_2(r_k)}} \right]^2
\end{equation}

\begin{equation}
\chi^2_{\langle \alpha \rangle,\sigma_{\alpha}}=  \sum_{k=1}^{N_r} \sum_{j=1}^2  \left[ \frac{M_j(r_k) - \langle M_j(r_k) \rangle}{\sigma_{M_j(r_k)}} \right]^2
\end{equation}

There is an ongoing discussion, whether a diagonal $\chi^2$-statistic or the ordinary $\chi^2$-statistic, which takes into account correlations among the different random variables through the covariance matrix is the better suited measure. On the one hand it is
important to take into account correlations among the test statistics,  on the other hand it has been argued by \citep{2004ApJ...612...64E} that the calculation of 
the  inverse covariance matrix may become numerically unstable when the correlations among the variables are strong making
the ordinary  $\chi^2$  statistic sensitive to fluctuations  rather than to absolute deviations.
For the WMAP 3-year data we follow the reasoning of \cite{2004ApJ...612...64E} and choose 
a diagonal $\chi^2$-statistics, because also in our case the moments are highly correlated 
leading to high values in the off-diagonal elements of the cross-correlation matrix. 
However, if the chosen model is a proper  description of the data, {\it any} combination of 
measures should yield statistically the same values for the observations and the simulations. 
Also for the \textit{a priori} scale-independent test statistics, where some unimportant scales contribute to the final 
value of $\chi^2$, we find significant signatures for non-Gaussianities in the northern sky. We detect non-Gaussianity for the full sky at a level of 96.9\% regarding the mean, 96.5\% for the standard deviation and 97.3\% for a combination of mean and standard deviation. For the northern hemisphere, the signatures of non-Gaussianities are more 
pronounced and we obtain 97.7\%  (mean), 99.5\%  (standard deviation) and 98.9\% (combination), whereas the Southern hemisphere is 
more consistent with Gaussianity [94.2\% (mean), 70.0\% (standard deviation) and 91.6\% (combination)]. These differences between 
Northern and Southern hemispheres induce pronounced asymmetries, which can be interpreted as a global lack of structure in the Northern hemisphere, 
which is consistent with previous findings (see below).

If we select only those pixels for $P(\alpha)$ which have $|b| > 30$ ($b$ galactic latitude), well outside the galactic plane, we get higher values for $P(\alpha)$. The disturbing edge effects of the Kp0-mask are almost totally removed. 
Only now we detect a localised anomaly in the Southern hemisphere when analysing the $\alpha$ spectra. We identify this signature as the Cold Spot, which was also already detected in the 
first-year WMAP data by \cite{2004ApJ...609...22V}. More local features are identified with the 5-year data and will be discussed in more detail in the following section. The probability densities $P(\alpha)$ of the selected pixels well outside the galactic plane are very similar to the former ones (see Figure \ref{fig5:probWMAP3}). The same holds for the significances for non-Gaussianity. 

\subsubsection{The WMAP five-year data}
\begin{figure}
\centering
\includegraphics[width=8cm, keepaspectratio=true, ]{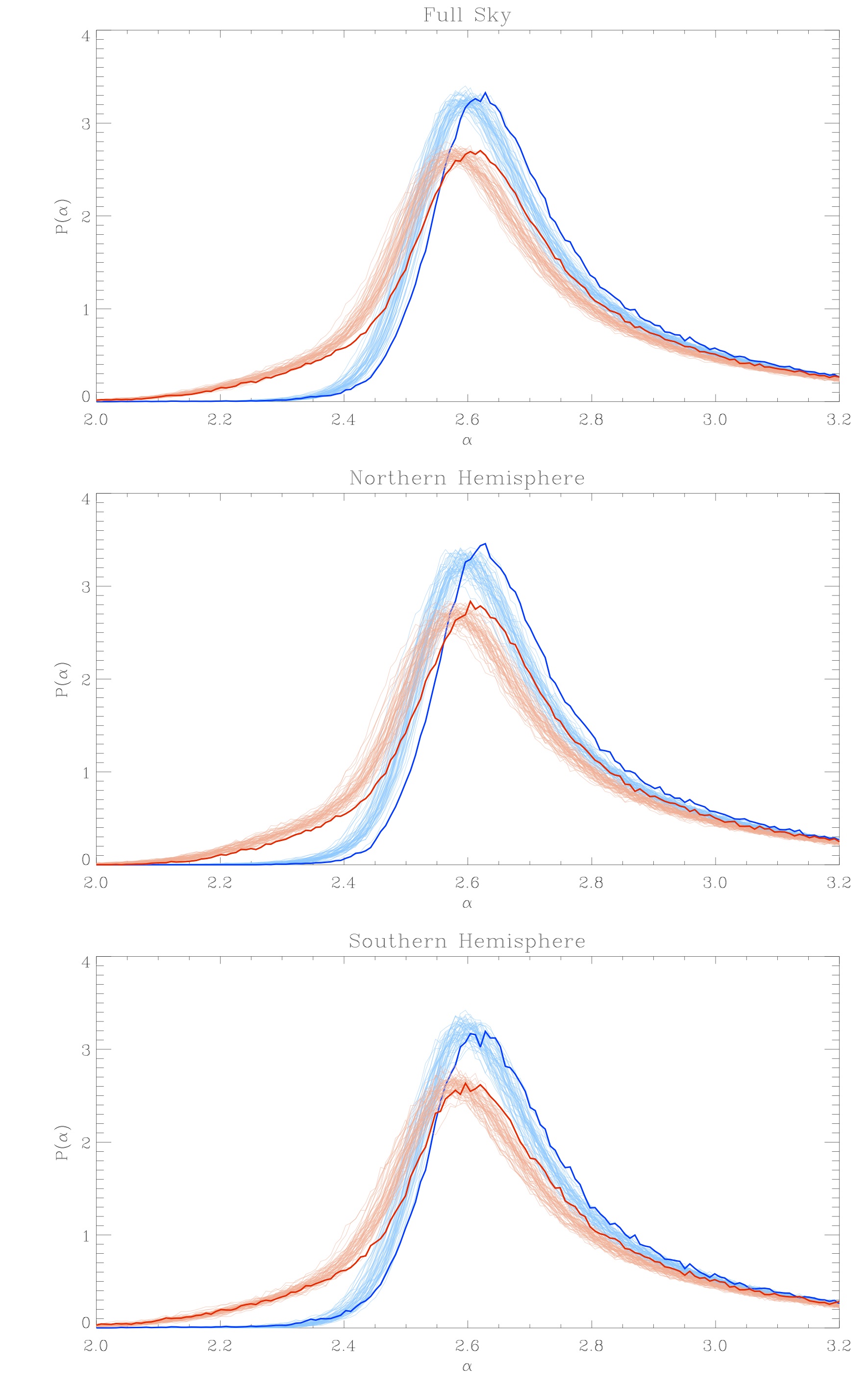}
\caption{The probability densities $P(\alpha)$ of the scaling indices for the WMAP 5-year data (dark lines) and for 50 simulations (fainter lines) by using the scale parameter $r=0.2$ and the KQ75-mask, computed for the original (red) and the mask filling method (blue). The upper histogram shows the distribution of the full sky data set, while the middle and the lower ones show the distribution of the northern and southern hemisphere, respectively.} \label{fig7:Overplots}
\end{figure}
The empirical probability densities $P(\alpha)$  of the scaling indices (calculated here with $r=0.2$) for WMAP 5-year data and respective simulations in Figure \ref{fig7:Overplots} show again a shift of the WMAP data to higher values, which becomes particularly apparent in the northern hemisphere of the galactic coordinate system. Comparing the non-filling and the mask filling method (see section \ref{WMAPi}), the histograms of the latter feature a higher maximum as well as higher values for large $\alpha$, but lower probabilities for $\alpha \in [2.0,2.5]$. The obvious reason for this shift is the fact that the filled mask does not reduce the $\alpha$-values of its surroundings as it was the case with the former method. Now, the outcome of these regions is influenced by the white noise and is therefore allocated at higher values.\par


\begin{figure*}
\centering
\includegraphics[width=17.4cm, keepaspectratio=true]{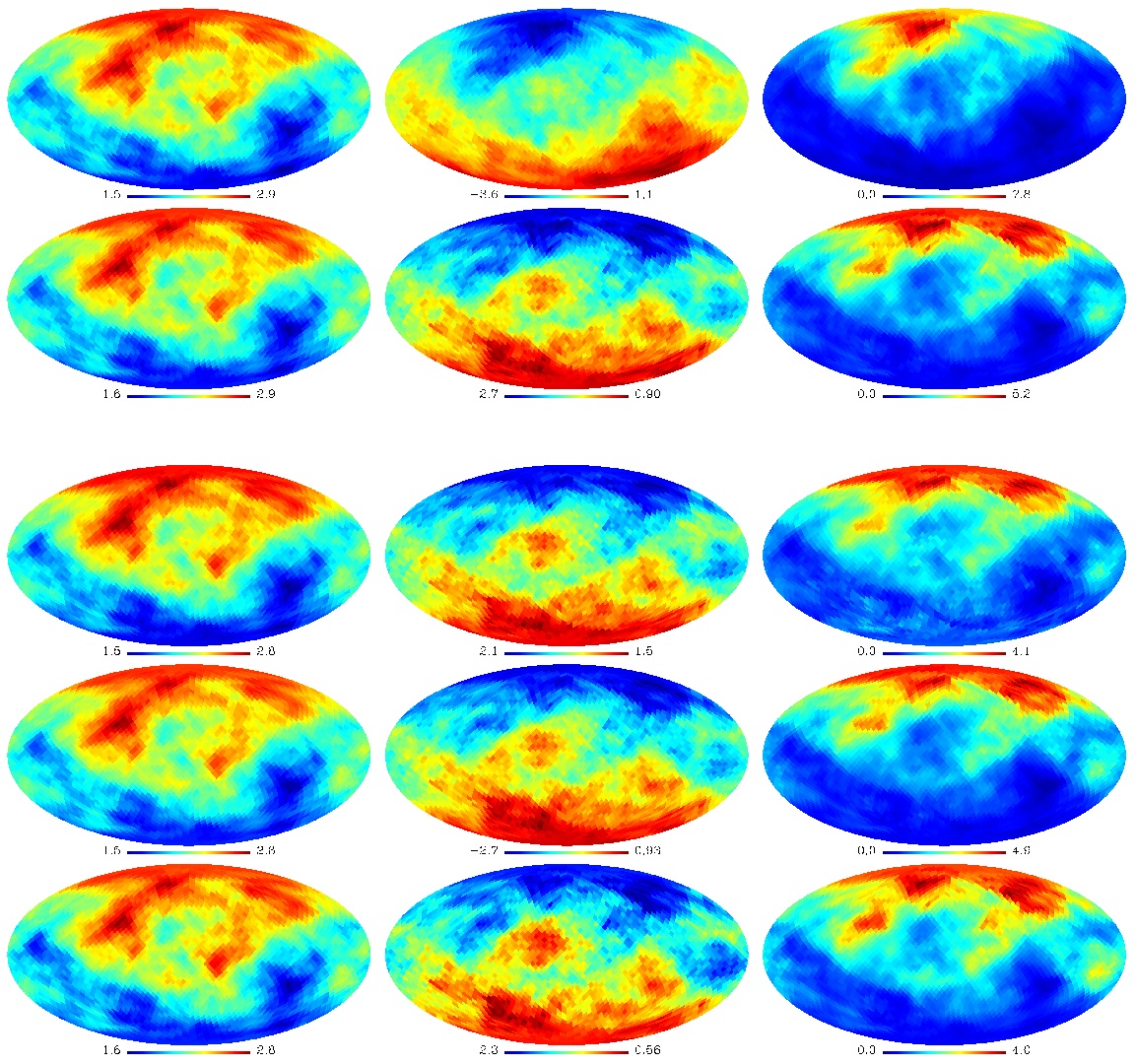}
\caption{The $\sigma$-normalised deviations $S(r)$ of the rotated hemispheres at the scale parameter $r=0.2$ for the mean (left column), the standard deviation (central column) and the diagonal $\chi^2$-statistics (right column) for the co-added VW-Band (WMAP 5-year) without (top row) and with (second row) the appliance of the mask filling method, as well as for the single Q-, V- and W-bands (third to fifth row), for which the mask filling method was always applied. Notice the different colour scaling for each plot.} \label{fig8:Rotationen}
\end{figure*}

We also calculated the $\sigma$-normalised deviations $S$ and the percentages $p$ of the simulations with higher (lower, respectively) results of the scale-independent diagonal $\chi^2$-statistics, again for $N=1000$ simulations. High deviations are found, particularly in the northern hemisphere. We derive evidence for non-Gaussianity with a probability of up to 97.3\% for the mean when regarding the KQ75-masked full sky and summing up over all considered length scales by means of a diagonal $\chi^2$-statistics. Looking at only the northern or southern hemisphere of the galactic coordinate system, we obtain up to 98.5\% or 96.6\%, respectively. For the standard deviation, these results appear as 95.6\% for the full sky (99.7\% north, 89.4\% south) and for a $\chi^2$-combination of both measurements as 97.4\% (99.1\% north, 95.5\% south). We obtain larger deviations from Gaussianity when looking at separate scale lengths. In general, all occurring characteristics match the findings of the analysis of the WMAP 3-year data. This indicates that the results are not based on some time-dependent effects. Since the 5-year data features lower error bars than the 3-year data, it is also improbable that both results are induced by noise effects only. \par

Evidence for north-south asymmetry in the WMAP data was already detected using the angular power spectrum \cite{2004MNRAS.354..641H, 2009ApJ...704.1448H} and higher order correlation functions \cite{2004ApJ...605...14E} spherical wavelets\cite{2004ApJ...609...22V}, local curvature analysis \cite{2004ApJ...607L..67H}, two-dimensional genus measurements \cite{2004MNRAS.349..313P} as well as all three Minkowski functionals \cite{2004ApJ...612...64E}, correlated component analysis \cite{2007MNRAS.382.1791B}, spherical needlets \cite{2008PhRvD..78j3504P}, frequentist analysis of the bispectrum \cite{2005MNRAS.357..994L}, two-point correlation functions \cite{2008MNRAS.389.1453B, 2008PhRvD..78f3531B} and Bayesian analysis of the dipole modulated signal model \cite{2009ApJ...699..985H}. To take a closer look at asymmetries in the WMAP 5-year data in our investigations, we perform an analysis of \textit{rotated} hemispheres and detect an obvious asymmetry in the data: For each scale we calculate the mean $\langle \alpha(r_k) \rangle$ and standard deviation $\sigma_{\alpha (r_k)}$ of the map of scaling indices $\alpha(\vec{p}_i,r)$ (or $\alpha(\theta, \phi; r_k)$) of the full sky and a set of 3072 rotated hemispheres. Every pixel centre of the full sky consisting of 3072 pixels ($N_{side} = 16$) in the HEALpix \cite{2005ApJ...622..759G} pixelisation scheme marks a new northern pole of one of the different hemispheres. The differences between the results of the original as well as the simulated maps are again quantified by the $\sigma$-normalised deviation $S$ (see equation \ref{eq:significance}). Every hemisphere of the set of $3072$ hemispheres delivers one deviation value $S$, which is then plotted on a sky map at the northern pole of the respective hemisphere. 

Thus, the colour of each pixel in the corresponding Figure \ref{fig8:Rotationen} expresses the positive or negative $\sigma$-normalised deviation $S(r)$ of the hemisphere around that pixel in the WMAP-data compared to the hemispheres around that pixel in the simulations. We apply this analysis for the co-added VW-band as well as for the single bands Q, V and W, whereas for the VW-band we use both the original and the mask filling method, but for the single bands the filling method only. In all charts of Figure \ref{fig8:Rotationen} we can detect an obvious asymmetry in the data: The largest deviations between the data and the simulations are exclusively obtained for rotations pointing to northern directions relative to the galactic coordinate system. The maximum value for $S(r)$ of the $\chi^2$ analysis (right column of Figure \ref{fig8:Rotationen}) using the mask-filling method on the co-added VW-band is obtained in the reference frame pointing to $(\theta,\phi) = (27^\circ,35^\circ)$, which is close to the galactic north pole. This proximity to the pole is consistent to the results of  \cite{2004ApJ...607L..67H} and \cite{2007MNRAS.380..466R}, as well as to those findings of  \cite{2004MNRAS.354..641H} and \cite{2004ApJ...605...14E} that consider large angular scales. For the standard deviation (central column of Figure \ref{fig8:Rotationen}), the northern and southern hemispheres offer different algebraic signs. The negative $S(r)$ of the north implies a lower variability than the simulations in this region, while the south shows a converse behaviour. The fact that the plots using the new method show slightly lower values for $S(r)$ than the ones using the old method may be explained by the fraction of pure noise values within every rotated hemisphere, that diminish the degree of difference between the data and the simulations. \par
While the Q-band is heavily foreground-affected, first of all by synchrotron radiation as well as radiation from electron-ion scattering ("free-free emission"), the W-band is mainly distorted by dust emission. The V-band is affected by all three of these foregrounds, even though less than the other bands. Despite the different influences on the different bands, we obtain the same signatures of non-Gaussianity in all single bands as well as in the co-added VW-band. The correlations $c$ of the different bands are high ($c \geq 0.95$). Therefore, we conclude that the measured asymmetry is not the result of a foreground influence but has to be concluded of thermal origin.
\par
\par
\par
An interesting anomaly in the CMB data is that there are small regions which show very high or very low values in some local structure analysis. One of the first of these local features, the well-known \textit{Cold Spot} at $(\theta,\phi) = (147^\circ,209^\circ)$, was first detected by \cite{2004ApJ...609...22V} in 2004 by using a wavelet analysis. Scaling indices were able to redetect the Cold Spot in the WMAP 3-year data (see section \ref{ch:wmap3}). Furthermore, it was identified using using amongst others wavelet analysis \cite{2004ApJ...613...51M,2005MNRAS.362..826C, 2005MNRAS.356...29C, 2007Sci...318.1612C} or the Kolmogorov stochasticity parameter \cite{2008A&A...492L..33G}. Furthermore, there have been some 
investigations which, in addition to the re-detection of the first spot, detected secondary spots via directional \cite{2005MNRAS.359.1583M,2006MNRAS.371L..50M,2008MNRAS.388..659M} or steerable wavelets \cite
{2007MNRAS.381..932V}, needlets \cite{2008PhRvD..78j3504P} and again the Kolmogorov stochasticity parameter \cite{2009A&A...497..343G}. These spots could be the result of some yet not fully understood physical 
process. For the Cold Spot lots of theories already exist which try to explain its origin by second-order gravitational effects \cite{2005PhRvD..72j3506T,2008PhRvD..77j3522T}, a finite universe model \cite
{2006gr.qc.....2102A}, large dust-filled voids \cite{2006ApJ...648...23I,2007ApJ...664..650I,2007ApJ...671...40R,2008ApJ...683L..99G}, cosmic textures \cite{2007Sci...318.1612C}, non-Gaussian modulation \cite{2010AstBu..65..101N}, topological defects \cite{2008JCAP...09..020B}, textures in a brane world model \cite{2008JCAP...10..039C} or an asymptotically flat Lemaître-Tolman-Bondi model 
\cite{2008JCAP...04..003G, 2009JCAP...07..035M}. 

\begin{figure*}
\centering
\includegraphics[width=8cm, keepaspectratio=true]{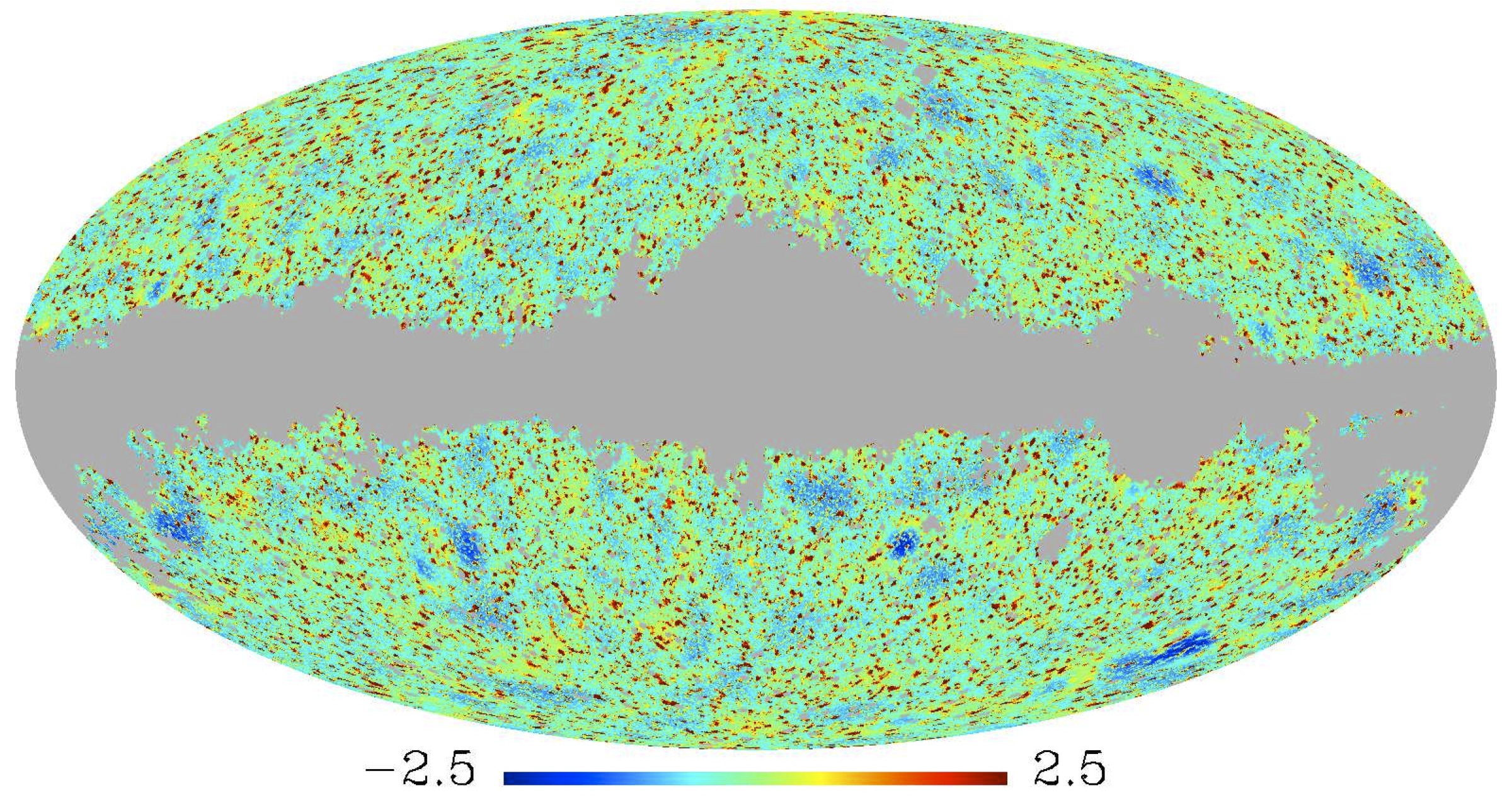}
\includegraphics[width=8cm, keepaspectratio=true]{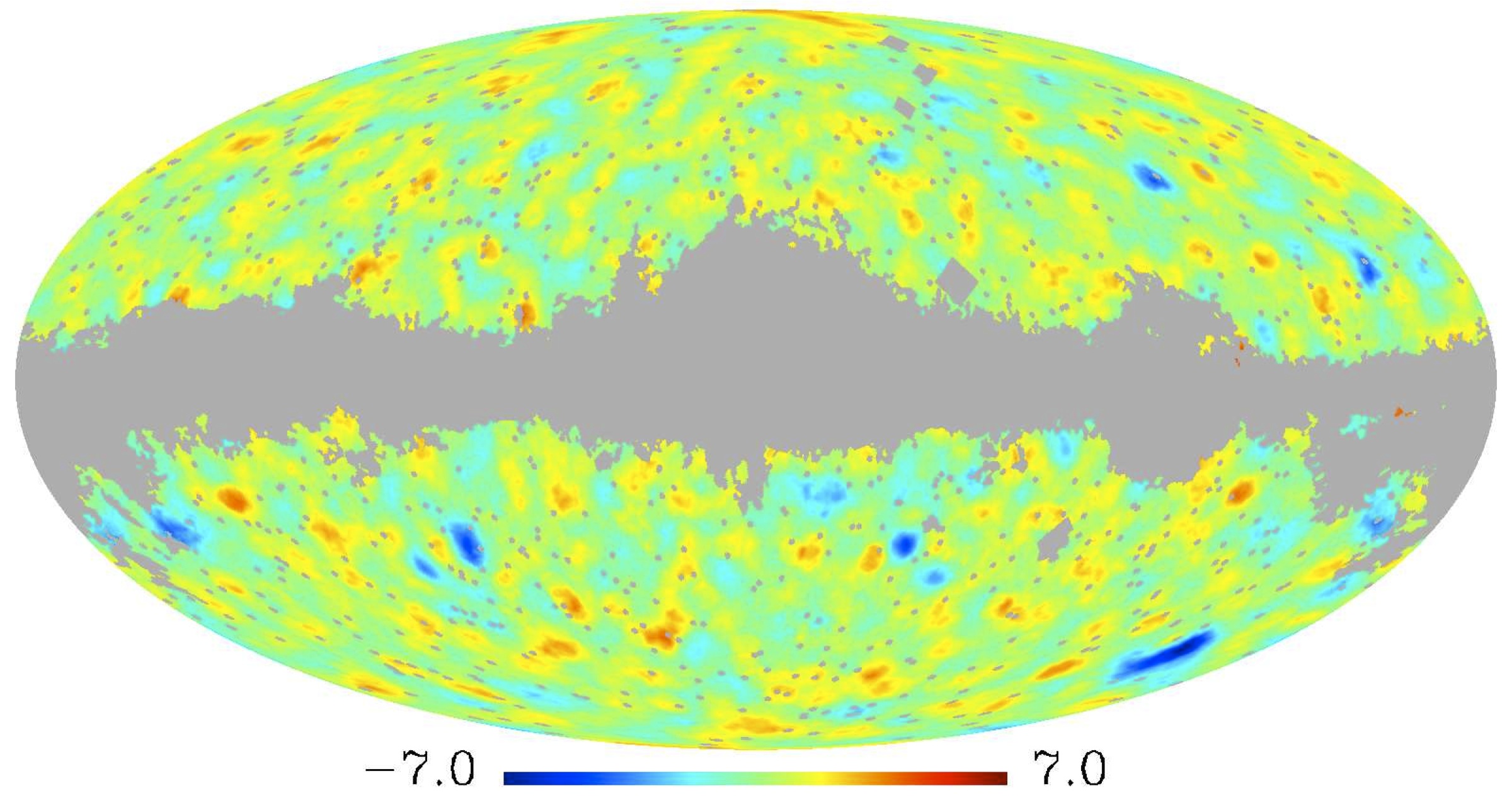}
\includegraphics[width=8cm, keepaspectratio=true]{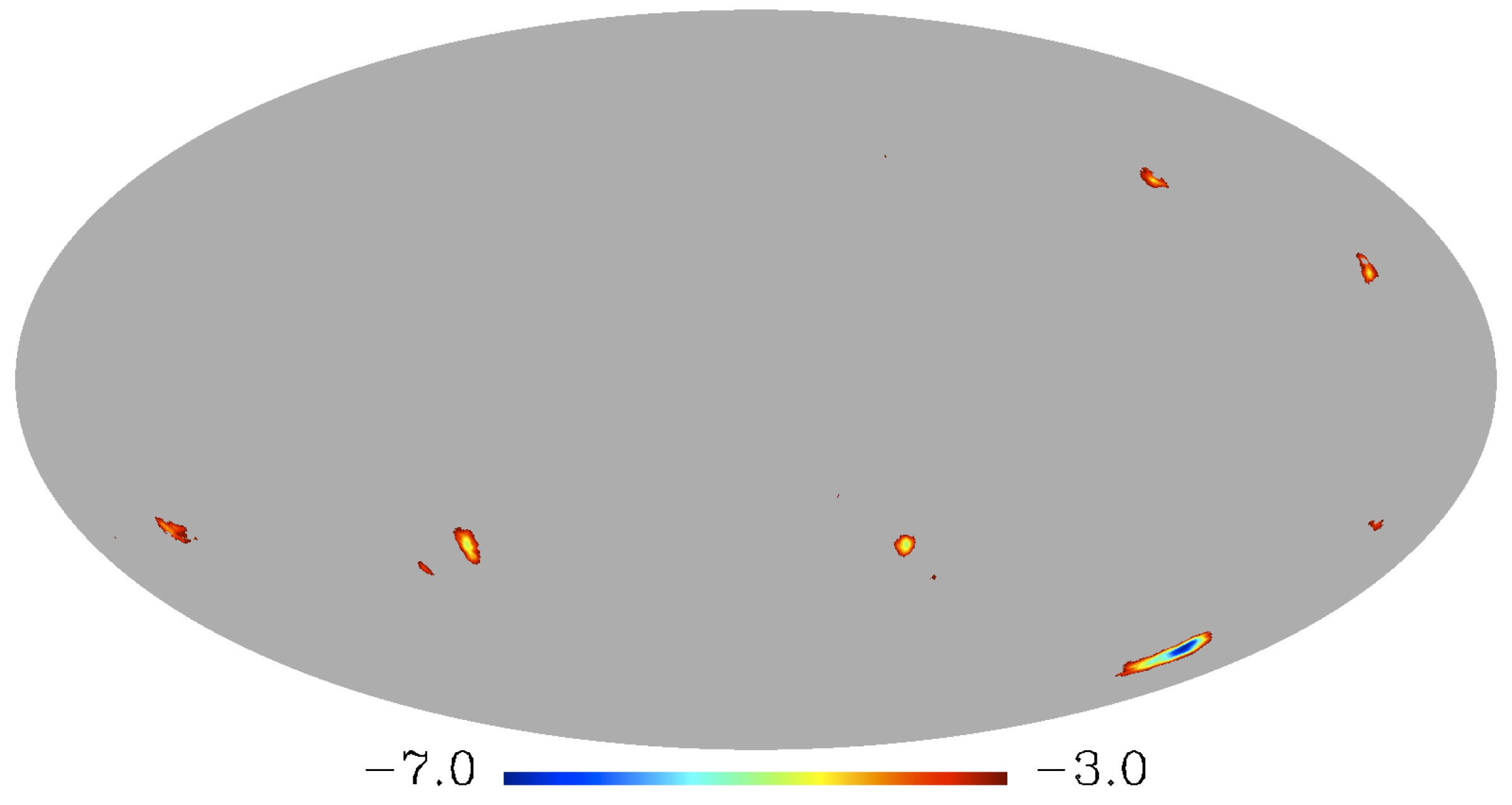}
\includegraphics[width=8cm, keepaspectratio=true]{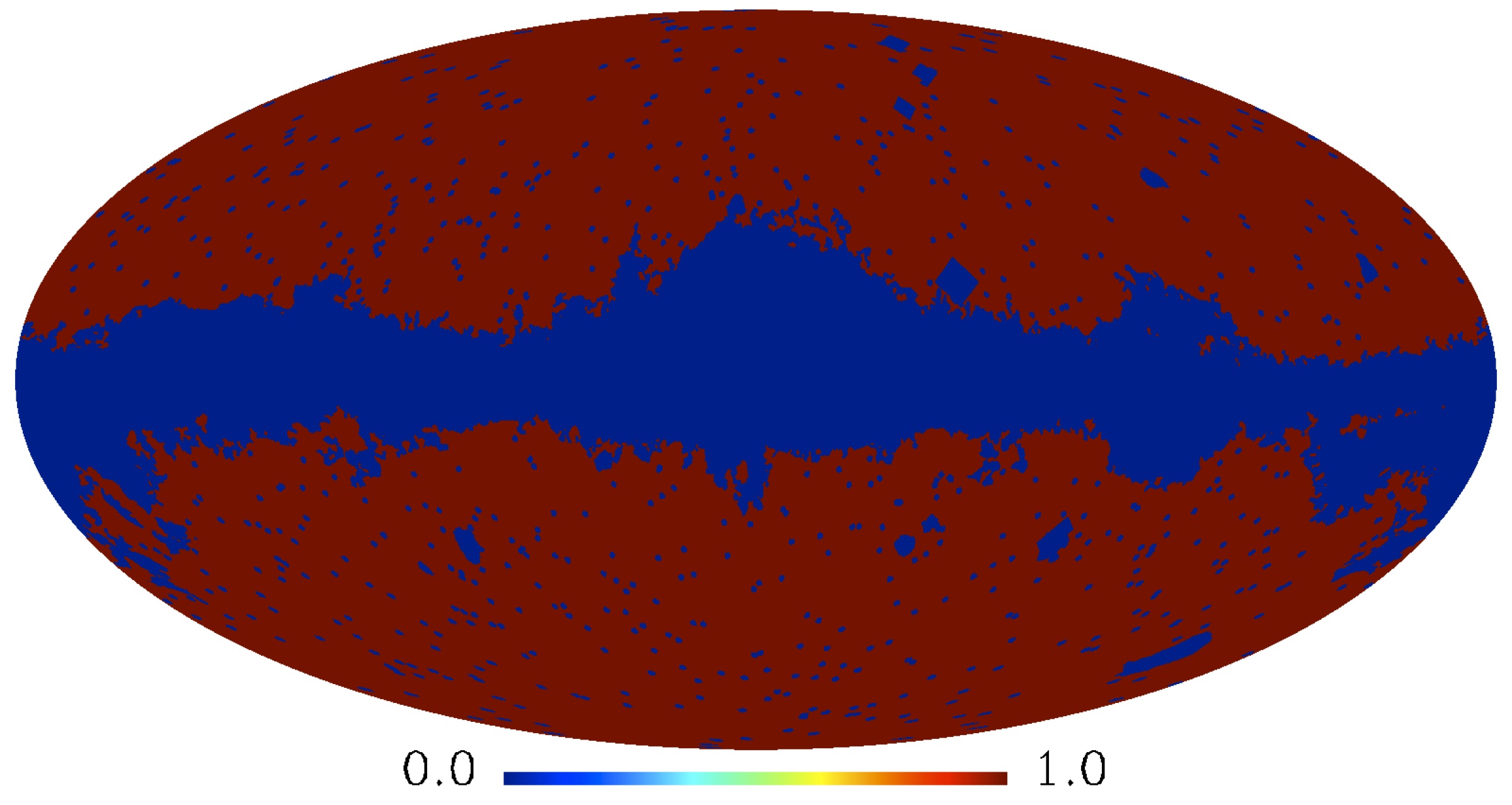}
\caption{The pixel-wise deviations $S_{i,r}$ of the primal (upper left) and of the smoothed scaling indices map (upper right), both based on the VW-band (WMAP 5-year) and the scale parameter $r=0.2$. The plot in the lower left only shows the values $\leq -3.0$ of the smoothed method. Except for the very small spots in the right part of this mapping, these regions are added to the KQ75-mask. The result is illustrated in the lower right plot.} \label{fig9:ColdSpots}
\end{figure*}

For our investigations concerning spots in the WMAP data we use the mask-filling method of section \ref{ch:maskfill}. Boundary effects caused by the mask are eliminated, which allows hidden effects to emerge. We extend the analysis of scaling indices by applying two different approaches to detect anomalies: The first one is to calculate the $\sigma$-normalised deviation of \textit{every pixel} on the $\alpha$-response of the CMB map. For a given scale parameter $r$, this is achieved by comparing the scaling index $\alpha(\vec{p}_i,r)$ of each vector $\vec{p}_i, i=1,...,N_{pix}$, of the original data with the mean of the corresponding values $\alpha_{\ell}(\vec{p}_{i},r), \ell = 1,...,N_{sim}$, of the simulations depending on their standard deviation, where $N_{sim}$ denotes the number of the simulations. Formally, this reads as:
\begin{equation} \label{PixelwSignifikFormel}
S_{i,r} = \frac{\alpha(\vec{p}_i,r) - \mu_{i,r}}{\sigma_{i,r}},
\end{equation}
with
\begin{align*}
\mu_{i,r} &= \frac{1}{N_{sim}} \sum_{\ell=1}^{N_{sim}} \alpha_{\ell}(\vec{p}_{i},r) \\
\sigma^2_{i,r} &= \frac{1}{N_{sim}-1} \sum_{\ell=1}^{N_{sim}} \left( \alpha_{\ell}(\vec{p}_{i},r) - \mu_{i,r} \right)^2
\end{align*}
The results are illustrated in the upper left part of Figure \ref{fig9:ColdSpots}. \par

\begin{figure}
\centering
\includegraphics[width=8cm, keepaspectratio=true]{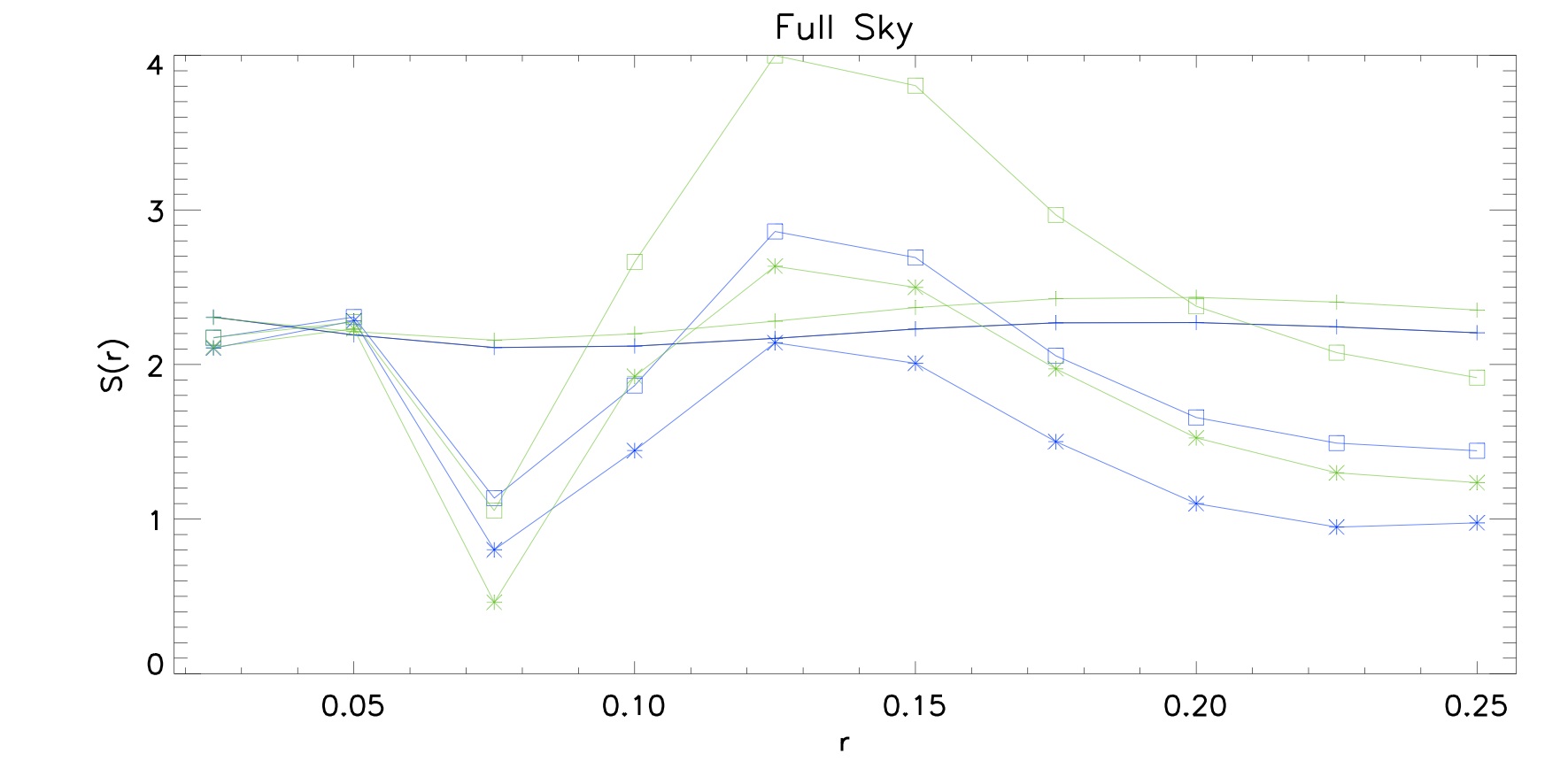}
\includegraphics[width=8cm, keepaspectratio=true]{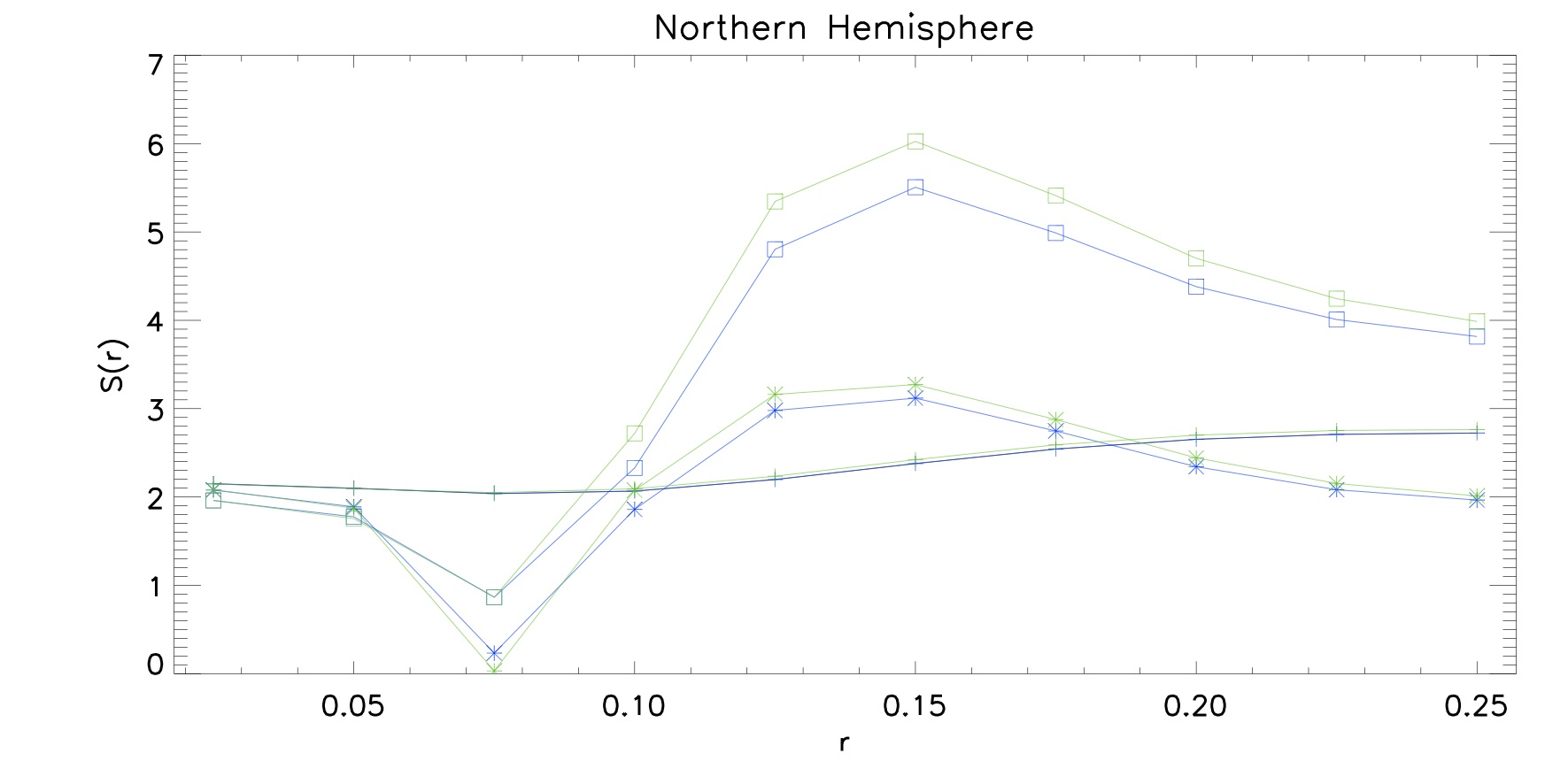}
\includegraphics[width=8cm, keepaspectratio=true]{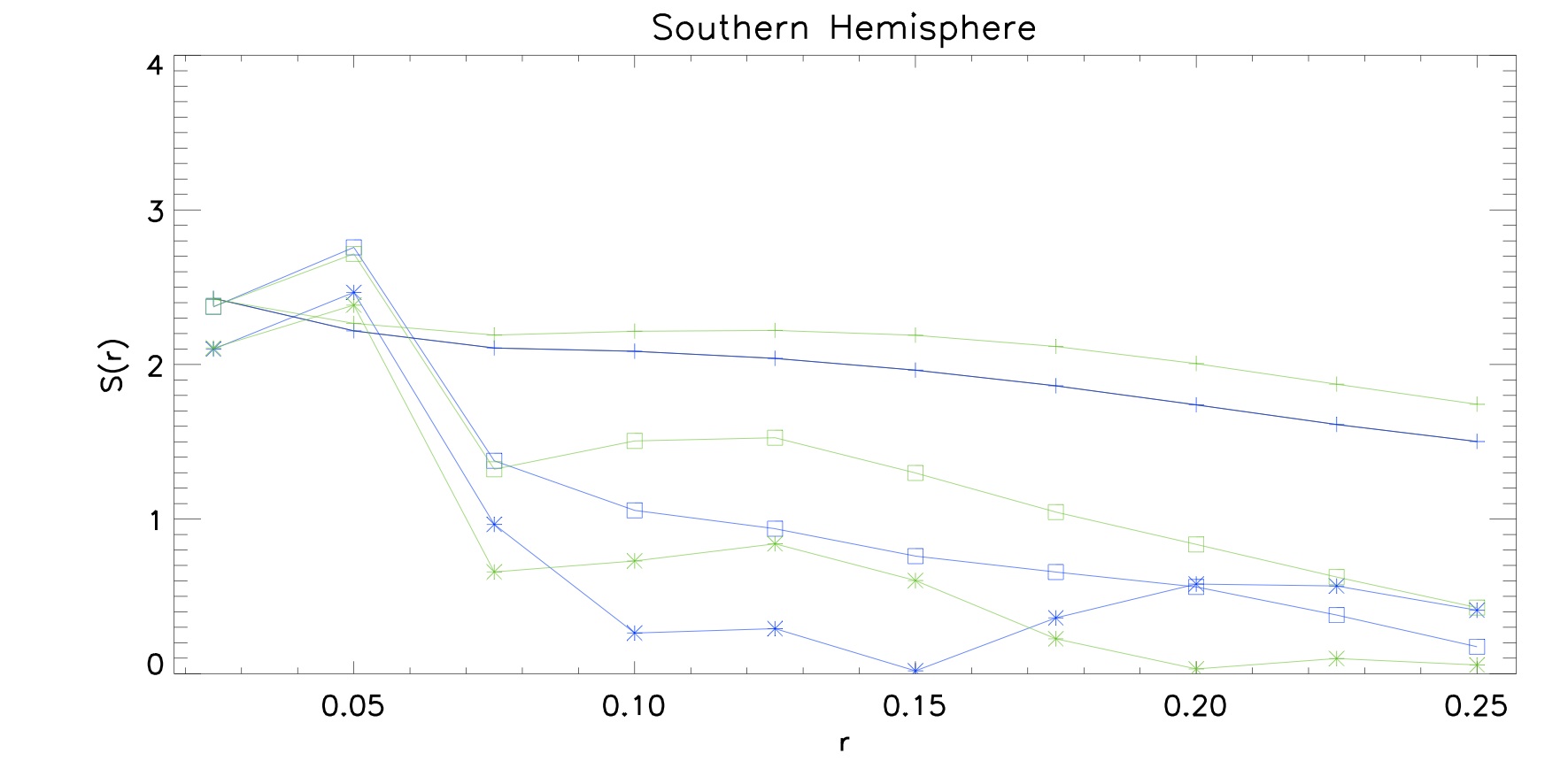}
\caption{The $\sigma$-normalised deviations of the mask-filling method for the original KQ75-mask (blue) and for the modified mask of the previous figure (green) in absolute values, plotted as a function of the scale parameter, whereby as above "$+$" denotes the mean, "$\ast$" the standard deviation and the boxes the $\chi^2$-combination. The full sky as well as again the single hemispheres are considered.} \label{fig10:SigVWnachMaske}
\end{figure}

The second approach smoothes the $\alpha$-maps of the original and simulated data by computing for every pixel the mean value of its surroundings given by some specified maximum distance, which equals $3^\circ$ in our analysis. We apply the pixel-wise deviations $S_{i,r}$ again on the resulting maps. The outcome of this procedure is shown in the upper right part of Figure \ref{fig9:ColdSpots}. In the lower left plot of the same Figure only the deviations $S_{i,r} \leq -3.0$ are illustrated to gain yet another clearer view on the interesting areas. We identify several local features on the map.\par
The first approach clearly shows the Cold Spot and indicates some secondary spots in the southern as well as in the northern hemisphere. These get confirmed in the plot of the smoothing method, where we obtain a deviation of up to $-7\sigma$ for several clearly visible areas: In the southern hemisphere we detect a cold spot at $(\theta,\phi) = (124^\circ,320^\circ)$ and another one at $(\theta,\phi) = (124^\circ,78^\circ)$. Both were already detected with the above mentioned directional and steerable wavelet as well as with a needlet analysis. The former one is a \textit{hot} spot in these investigations. In our analysis, the latter spot actually appears as two spots close to each other, which is in agreement with \cite{2008PhRvD..78j3504P}. We discover another southern cold spot at $(\theta,\phi) = (120^\circ,155^\circ)$ which is very close to the mask. This spot represents a good example for the use of the mask filling method since it is located at the edge of the non-masked region: The influence of the mask is diminishing the results of the calculation of the scaling indices in the area of this spot. This becomes obvious if one recalls the lower left plot of Figure \ref{fig1:dreimaldrei}, in which the coordinates of the spot would be completely located in a "blue" region with low $\alpha$-values. Since the results of the scaling indices of local features show a similar, namely lower-valued, behaviour, an overlapping like that could prevent the detection of such spots close to the mask. By using the mask filling method, the detection of this cold spot on the edge of the mask is equivalent to a detection in an unmasked region, and therefore reliable. The spot at $(\theta,\phi) = (136^\circ,173^\circ)$, described by \cite{2005MNRAS.359.1583M} and \cite{2008PhRvD..78j3504P}, is not traced in our analysis. In the northern hemisphere, our investigation shows two other cold spots at $(\theta,\phi) = (49^\circ,245^\circ)$ and $(\theta,\phi) = (68^\circ,204^\circ)$, which do not correspond with the so-called \textit{northern cold spot} of \cite{2009A&A...497..343G}, but with the results of \cite{2005MNRAS.359.1583M}, where again one of them is a hot spot. Also \cite{2008PhRvD..78j3504P} locates one of these two spots. All these results were achieved with an analysis of the VW-band, but we find similar results in a single band analysis. \par
If the considered spots really depend on some yet not completely understood, maybe secondary, physical effect, they should not be implemented in a testing for intrinsic non-Gaussianity. For this reason, we modify the 5-year KQ75-mask by additionally excluding all above mentioned spots. A small peculiarity at the edge of the mask next to the Cold Spot as well as three very small blurs in the right half of the lower left Mollweide projection in Figure \ref{fig9:ColdSpots} are not considered, since we regard their appearance as insufficient for being a distinctive feature. The modification of the KQ75-mask is illustrated in the lower right part of Figure \ref{fig9:ColdSpots}. \par
We now apply this new mask to the $\alpha$-response of both the WMAP data as well as the simulations and repeat the analysis from above. When excluding all these spots from the analysis, the deviation from Gaussianity increases, which shows that the discovered local anomalies are not the reason of the global detection of non-Gaussianity, but actually were damping the deviations on average. The results of the $\sigma$-normalised deviations $S$ are illustrated in Figure \ref{fig10:SigVWnachMaske}. An increase of $S(r)$ in comparison to the former analysis where the usual KQ75-mask was used (results not shown, please refer to \cite{2007MNRAS.380..466R} ) is in particular present in the southern hemisphere, where we detected more local features than in the north. The largest increase takes place in the co-added VW-band, where we now reach deviations of up to 4.0 for the $\chi^2$-combination in a full-sky analysis (former maximum: 2.9) and to the extend of 6.0 in an analysis of the northern hemisphere (former maximum: 5.5). Also the single bands as well as all scale-independent diagonal $\chi^2$-statistics show without exception a greater evidence for non-Gaussianity. \par

\subsection{Comparison with Surrogates}

\begin{figure*}
\centering
\includegraphics[width=17.4cm, keepaspectratio=true]{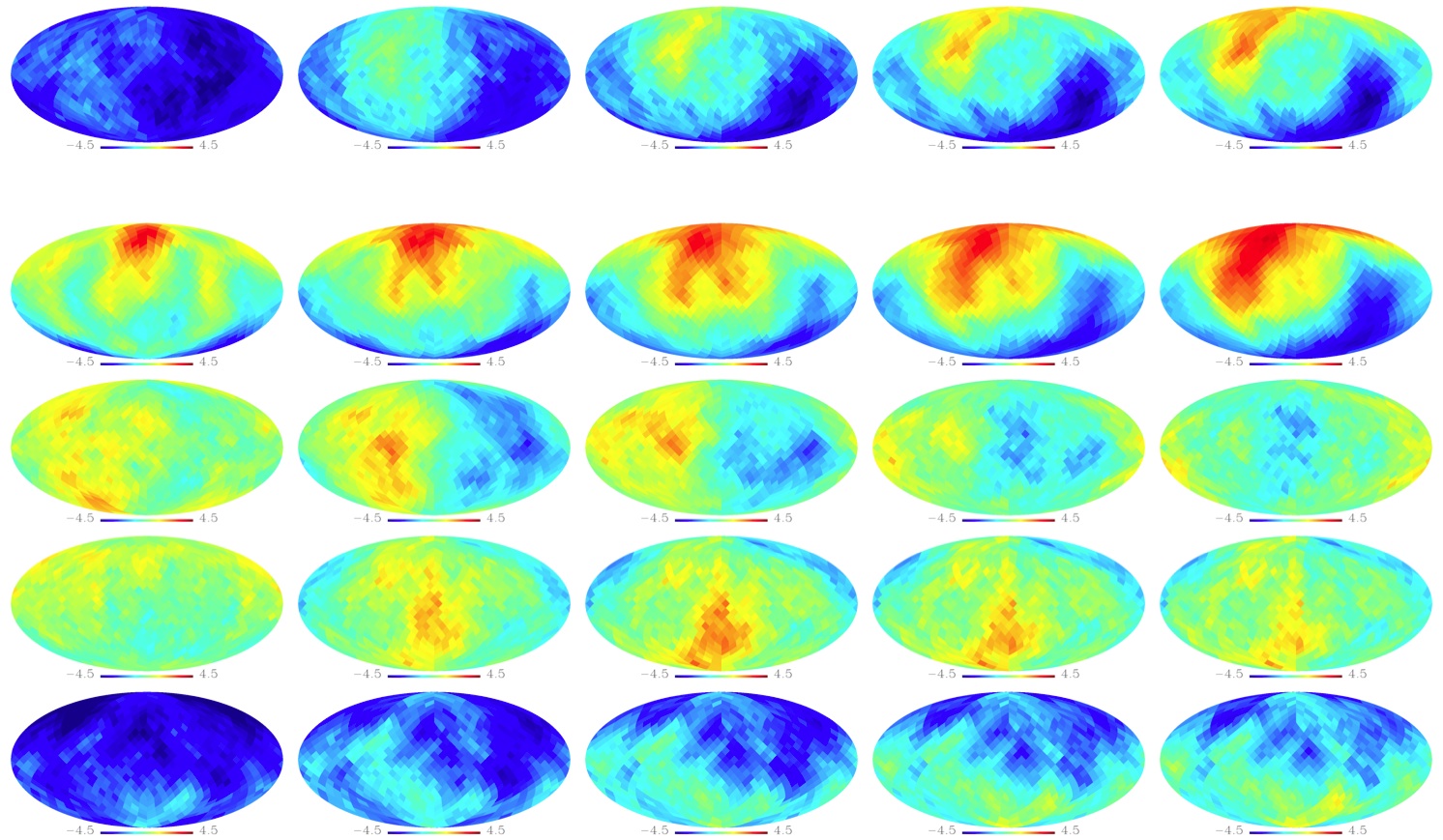}
\caption{Deviations $S(\langle \alpha (r_{k}) \rangle)$ of the rotated hemispheres for the five scales $r_{k}, k=2,4,\ldots,10$ 
(from left to right) for the ILC7 map and for $\Delta \ell = [2,1024]$, $\Delta \ell = [2,20]$, $\Delta \ell = [20,60]$, $\Delta \ell = [60,120]$ and
$\Delta \ell = [120,300]$ (from top to bottom).} \label{fig11:rotWMAP7}
\end{figure*}

\begin{figure*}
\centering
\includegraphics[width=17.4cm, keepaspectratio=true]{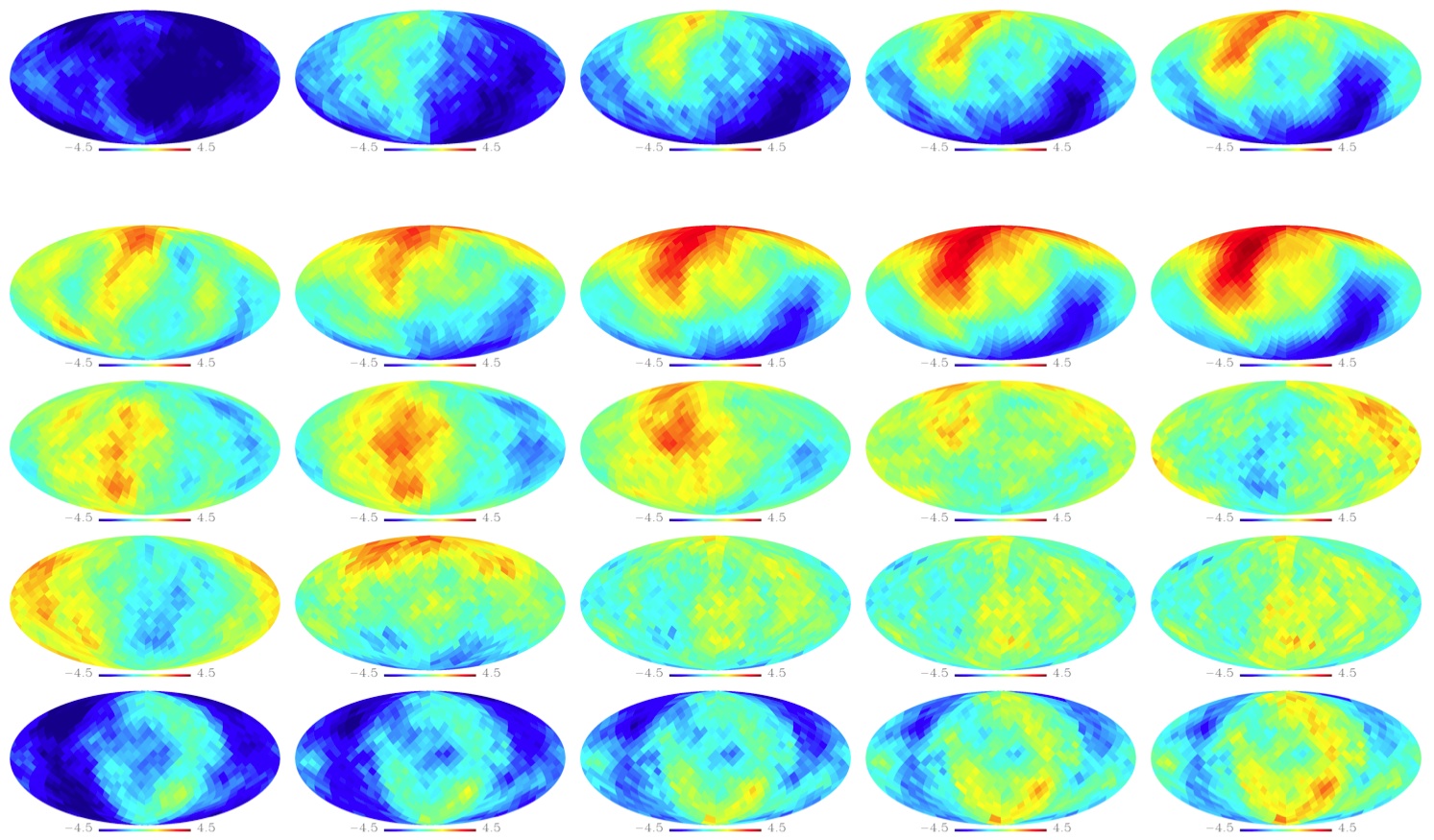}
\caption{Same as Figure \ref{fig11:rotWMAP7} but for the needlet-based NILC5 map.} \label{fig11a:rotNILC5}
\end{figure*}

We compare the first and second order surrogate maps by calculating the $\sigma$-normalised deviations $S$ (similar to  equation \ref{eq:significance}) 
\begin{equation}\label{eq:significance2}
S(Y) = \frac{Y_{surro1} - \langle Y_{surro2} \rangle}{\sigma_{Y_{surro2}}}
\end{equation}
between the two classes of surrogates for a set of (now) 768 hemispheres to test for NGs and asymmetries in the ILC7 map and the NILC5 map. Figure \ref{fig11:rotWMAP7} shows the deviations $S$ per hemisphere for the mean value $S(\langle \alpha(r_k) \rangle), k=2, 6, 10$ for the ILC7 map 
as derived from the comparison of the different classes of surrogates for the scale-independent surrogate test and for the four selected $\ell$-ranges.
The following striking features become immediately obvious:\\
First, various deviations representing features of non-Gaussianity and asymmetries can be found in the $S$-maps for the ILC7 map. 
These features can nearly exactly be reproduced when the NILC5 map is taken as input map, whose results are illustrated in Figure \ref{fig11a:rotNILC5}. \\
Second, we find for the scale-independent surrogate test (top rows in Figures \ref{fig11:rotWMAP7} and \ref{fig11a:rotNILC5}) 
large isotropic deviations for the scaling indices calculated for the small scale $r_2$. 
The negative values for $S$ indicate that the mean of the scaling indices for the first order surrogate is smaller than for the second order surrogates.
This systematic trend can be interpreted such that there is more structure detected in the first order surrogate 
than in the second order surrogate maps. Obviously, the random shuffle of all phases has destroyed a significant amount of 
structural information at small scales in the maps.\\ 
Third, for the scale-dependent analysis we obtain for the largest scales ($\Delta \ell = [2,20] $) (second lines in Figures \ref{fig11:rotWMAP7} and \ref{fig11a:rotNILC5}) highly significant signatures for 
non-Gaussianities and ecliptic hemispherical asymmetries at the largest $r-$values. 
These results are perfectly consistent  with those obtained for the  WMAP 5-year ILC map and the foreground 
removed maps generated by \cite{2003PhRvD..68l3523T} on the basis of the WMAP 3-year data (see \cite{2009PhRvL.102m1301R}). 
The only difference between this study and our previous one is that 
we now obtain higher absolute values for $S$ ranging now from $-4.00 < S < 3.72$ for the ILC7 map
and  $-4.36 < S < 4.50$ for  the NILC5 map as compared to $-3.87 < S < 3.51$ for the WMAP 5-year ILC map.
Thus, the cleaner the map becomes due to better signal-to-noise ratio and/or improved map making techniques the
higher the significances of the detected anomalies, which suggests that the signal is of intrinsic CMB origin.\\
Fourth,  we also find for the smallest considered scales ($\Delta \ell = [150,300] $)  
large isotropic deviations for the scaling indices calculated for a small scaling range $r$ very similar to those 
observed for the scale-independent test.\\
Fifth, we do not observe very significant anomalies for the two other bands 
($\Delta \ell = [20,60] $ and $\Delta \ell = [60,120] $) being considered in this study. 
Thus, the results obtained for the scale independent surrogate test can be interpreted
as a superposition of the signals identified in the two $\ell$-bands covering the largest ($\Delta \ell = [2,20] $) and 
smallest $\Delta \ell = [120,300] $) scales.

\begin{figure}
\centering
\includegraphics[width=8.5cm, keepaspectratio=true]{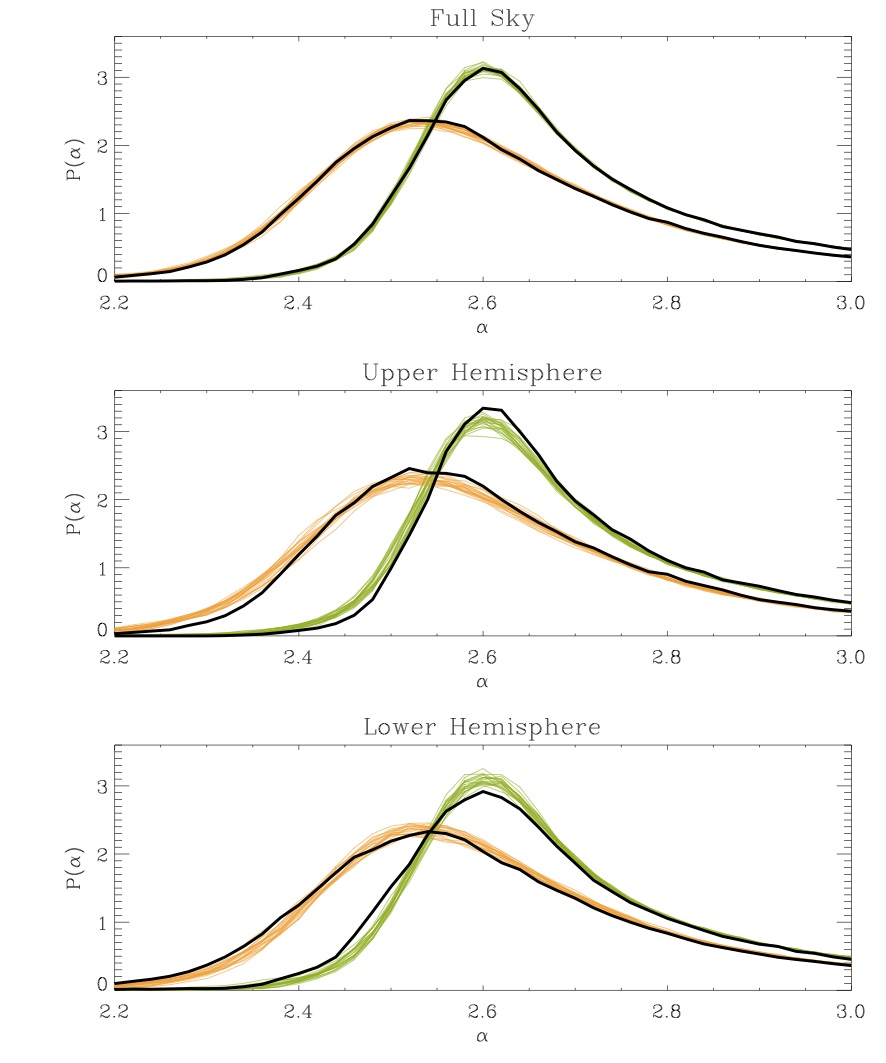}
\caption{The probability densities $P(\alpha)$ of the scaling indices for the first (black) and second order surrogates (coloured)
of WMAP 7-year data, calculated for the largest scaling range $r_{10}$ and
for the $\ell-$interval $\Delta \ell = [2,20]$. Yellow (green) curves denote the densities for $20$ realizations of 
second order surrogates derived from the  ILC7  (NILC5) map. The reference frame for defining the upper and lower hemispheres 
is chosen such that the difference $\Delta S = S_{up} - S_{low}$ becomes maximal for $\langle \alpha \rangle$
of the respective map and respective scale $r$.} \label{fig12:probWMAP7}
\end{figure}

\begin{figure}
\includegraphics[width=8.25cm, keepaspectratio=true]{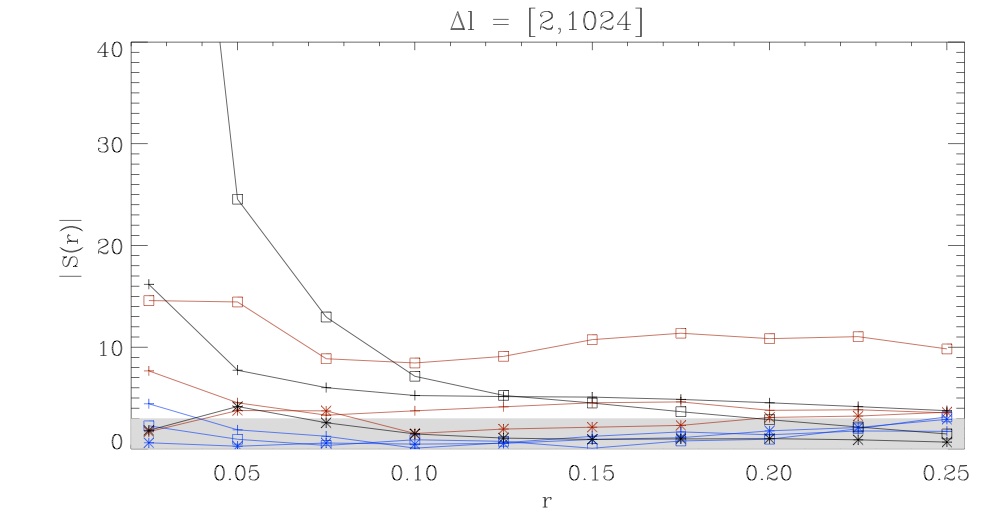}
\includegraphics[width=8.25cm, keepaspectratio=true]{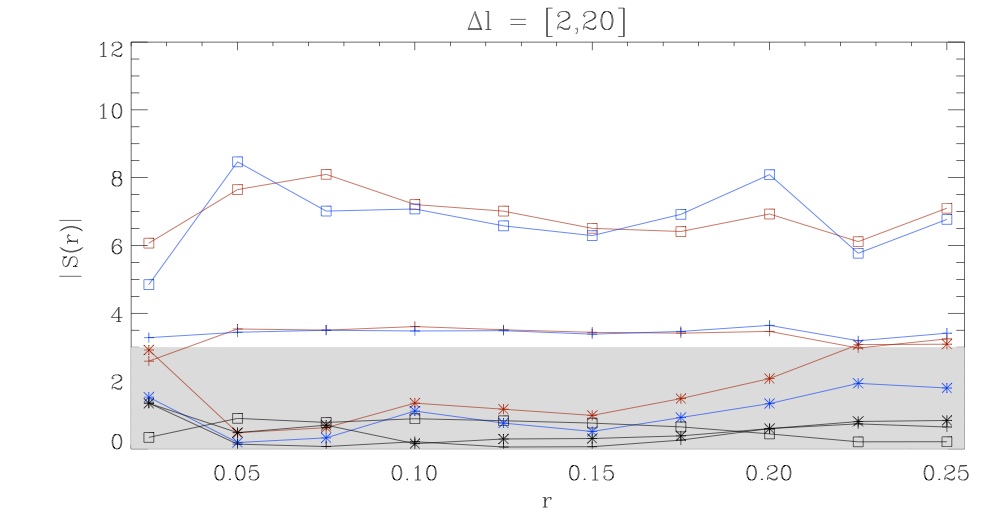}
\includegraphics[width=8.25cm, keepaspectratio=true]{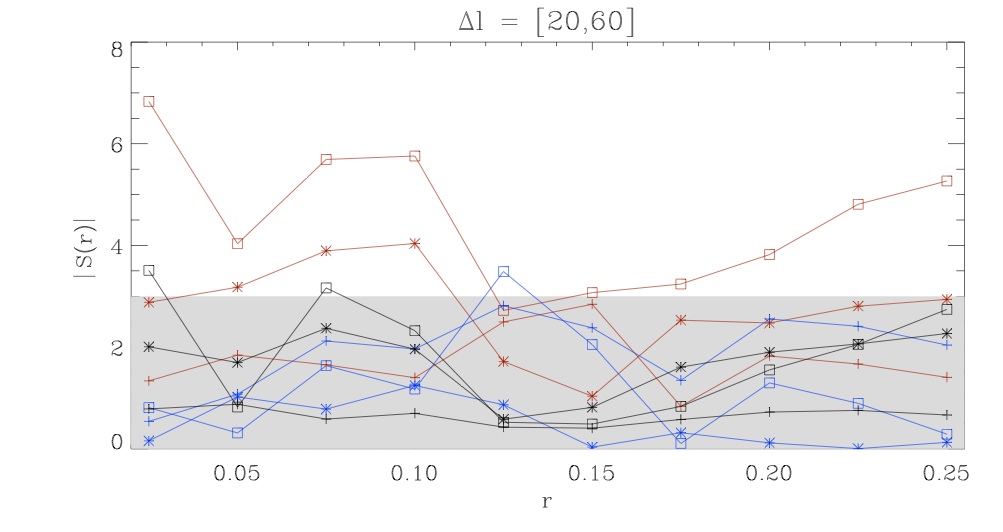}
\includegraphics[width=8.25cm, keepaspectratio=true]{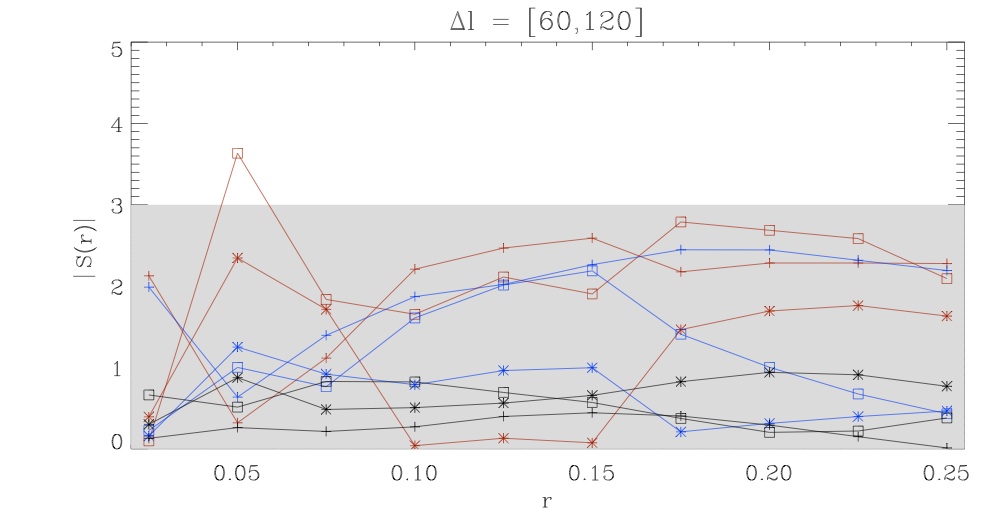}
\includegraphics[width=8.25cm, keepaspectratio=true]{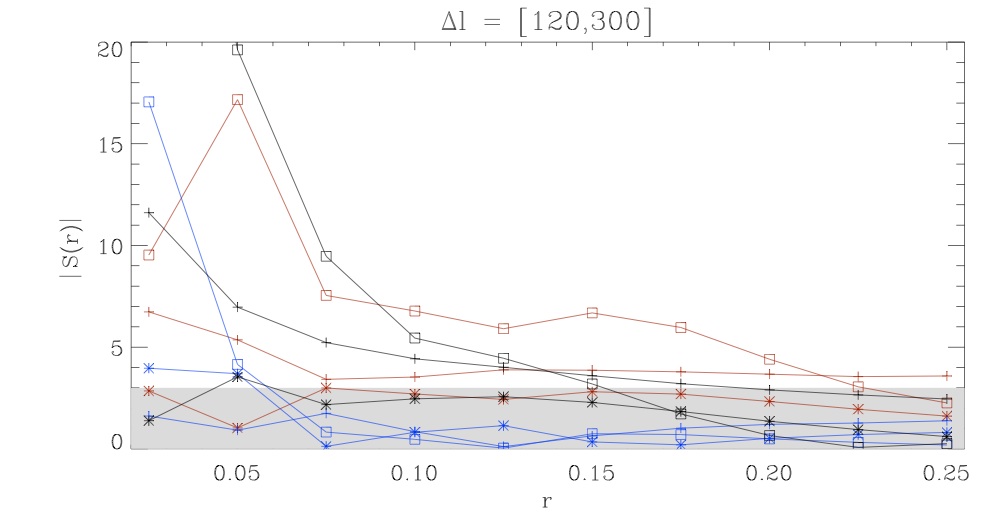}

\caption{Deviations $|S(r)|$ for the ILC7 map and the considered $\ell-$intervals  as a function of 
                the scale parameter $r$ for the full sky (black) and 
                the upper (red) and lower (blue) hemisphere. The '$+$' symbols denote 
                the results for the mean $\langle \alpha (r_{k}) \rangle$, the'$\ast$' symbols for the standard deviation  $\sigma_{\alpha (r_{k})} $
                and the boxes for the $\chi^2$-combination of $\langle \alpha (r_{k}) \rangle$ and $\sigma_{\alpha (r_{k})} $. } \label{fig13:devWMAP7}
\end{figure}

\begin{figure}
\includegraphics[width=8.25cm, keepaspectratio=true]{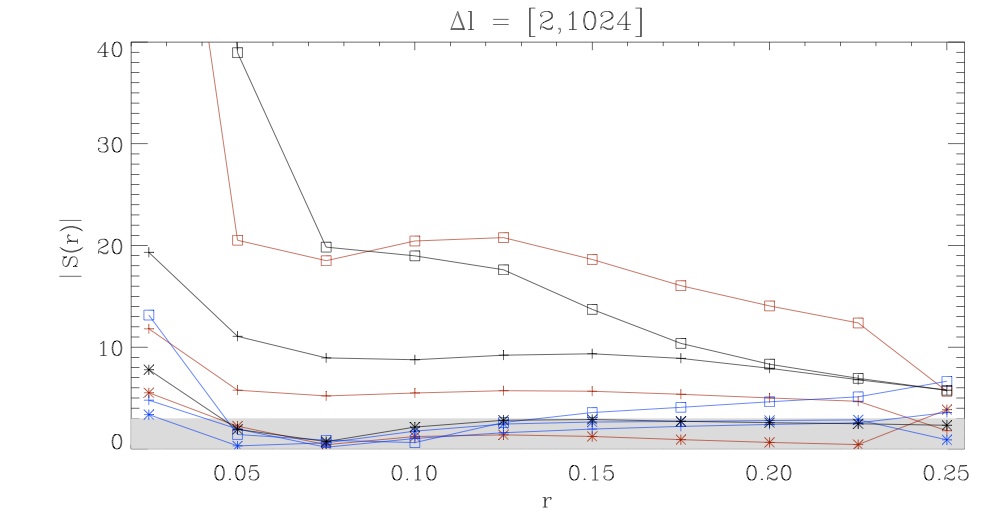}
\includegraphics[width=8.25cm, keepaspectratio=true]{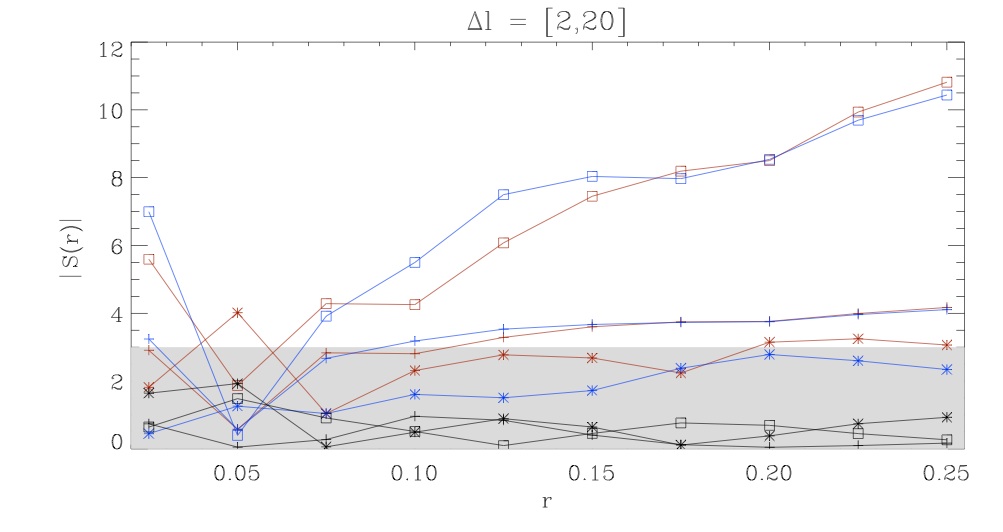}
\includegraphics[width=8.25cm, keepaspectratio=true]{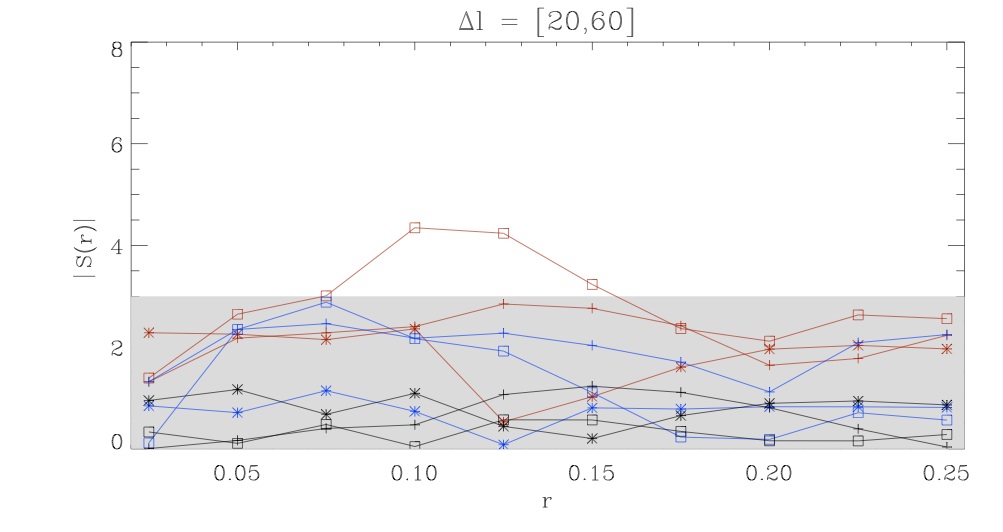}
\includegraphics[width=8.25cm, keepaspectratio=true]{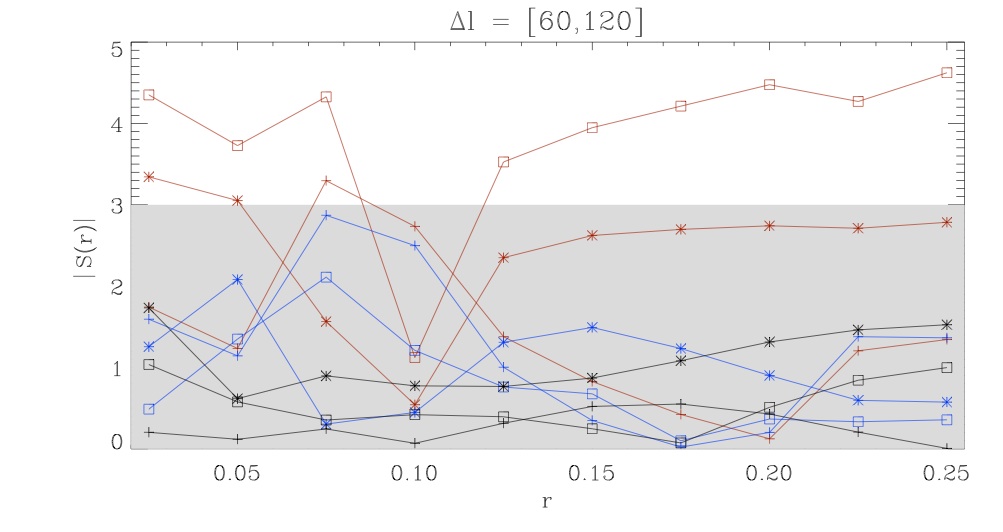}
\includegraphics[width=8.25cm, keepaspectratio=true]{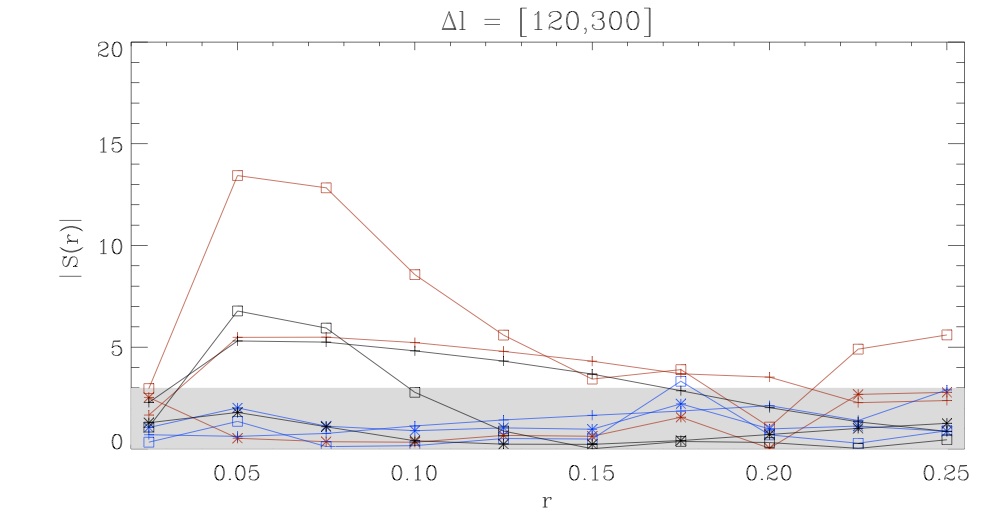}

\caption{Same as Figure \ref{fig13:devWMAP7}, but for the NILC5 map.} \label{fig14:devWMAP7nilc}
\end{figure}

Figure \ref{fig12:probWMAP7} shows the probability densities derived for the full sky and for (rotated) hemispheres
for the scaling indices at the largest scaling range $r_{10}$ for the 
first and second order surrogates for the $\ell$-interval $\Delta \ell = [2,20] $. We recognise the systematic shift of the whole density distribution 
towards higher values for the upper hemisphere and to lower values for the lower hemisphere.
As these two effects cancel each other for the full sky, we do no longer see significant differences
in the probability densities in this case.
Since the densities as a whole are shifted, the significant differences between first and second order surrogates 
found for the moments cannot be attributed to some salient localizable features leading to an excess 
(e.g. second peak) at very low or high values in otherwise very similar $P(\alpha)$-densities. Rather, the shift to 
higher (lower) values for the upper (lower) hemisphere must be interpreted as a global trend indicating that
the first  order surrogate map has less (more) structure than the respective set of second order surrogates.
The seemingly counterintuitive result for the upper hemisphere is on the other hand consistent with a linear 
hemispherical structure analysis by means of a power spectrum analysis, where also a lack of power in the 
northern hemisphere and thus a pronounced hemispherical asymmetry was detected \cite{2004ApJ...607L..67H, 2009ApJ...704.1448H}.
However, it has to be emphasised that the effects contained in the power spectrum are -- by construction -- exactly
preserved in both classes of surrogates, so that the scaling indices measure effects that can solely be induced by 
HOCs thus being of a new, namely non-Gaussian, nature. 
Interestingly though, the linear and nonlinear 
hemispherical asymmetries seem to be correlated with each other.\\
The density distributions derived from the ILC7 and NILC5 map are clearly shifted against each other. The differences between these two maps can be attributed to e.g. the smoothing of the ILC7 map. 
However, the systematic differences between first and second order surrogates induced by the phase manipulations
prevail in all cases -- irrespective of the input map.\\
The results for the deviations $|S(r)|$ for the full sky and rotated upper and lower hemisphere 
are shown for all considered $\ell$-ranges and all scales $r$ in Figures \ref{fig13:devWMAP7} and \ref{fig14:devWMAP7nilc}. Using scale-independent $\chi^2_{\langle \alpha \rangle,\sigma_{\alpha}}$-statistics combining the mean and the standard deviation and summing up over all scales $r$, the largest values for $S$ are found for the largest $ \Delta \ell = [2,20]$ and smallest scales $ \Delta \ell = [120,300]$ and for the scale-independent NGs. For the full sky non-Gaussianity is detected with a probability of up to 94.2\% ($\chi^2_{\langle \alpha \rangle,\sigma_{\alpha}}$) and 99.8\% for the northern and southern hemispheres. \par
To test whether all these signatures are of intrinsic cosmic origin or more likely due to foregrounds or 
systematics induced by e.g. asymmetric beams 
or map making, we performed the same surrogate and scaling indices analysis for five additional maps 
described in \cite{2010arXiv1012.2985R}. This set of tests to investigate whether and to what extend the detected anomalies 
can be explained by systematics cannot convincingly rule out the intrinsic nature of the anomalies for the
low $\ell$ case, while the ILC map making procedure and/or residual noise in the maps can also lead 
to NGs at small scales.

\section{Conclusions}\label{Conclusions}

In this \textit{Review}, we gave an overview of the application of scaling indices in CMB analysis to date. The SIM is a measure that detects different forms of topological behaviour in the data, which turned out to be very useful for identifying deviations from Gaussianity and statistical isotropy in the spherical data set of the microwave background radiation. In the following, the large number of deviations from Gaussianity and statistical isotropy detected by means of the SIM is summarised and the resulting conclusions are drawn.

By comparing the 3-year and 5-year measurements of the WMAP satellite with simulated CMB maps, several clear non-Gaussianities as well as asymmetries were detected:

The spectrum of scaling indices of the data is systematically broader and shifted towards higher values than the one of the simulations, yielding highly significant deviations of the mean, the standard deviation and a $\chi^2$-combination. These effects can be interpreted as too few structure and structural variations in the temperature anisotropies as measured by WMAP compared to the predicted ones within the concordance model, which is in agreement with previous results (e.g. \cite{2004ApJ...605...14E, 2004ApJ...612...64E, 2005PhRvL..95g1301L, 2009ApJ...704.1448H, 2009ApJ...699..985H}). By performing an analysis of rotated hemispheres, the rotations pointing to northern directions show by far higher deviations from Gaussianity for the mean and the $\chi^2$ analysis than rotations pointing to the south. For the standard deviation, the rotated hemispheres show a negative outcome in the north and a positive in the south. This implies that the north possesses a more consistent pattern than the simulations, while the south shows a converse behaviour.

All these results are consistent in different ways: Since the detected effects are the same for the 3-year as well as for the 5-year WMAP data, they can be concluded to be time-independent. In addition, the findings are nearly the same for the different bands that were analysed for the 5-year data, which leads to the conclusion that the foreground influence only plays a minor role. Furthermore, the usage of the mask-filling method, again applied to the 5-year data only, reduces the distorting influence of the mask. Since this leads to similar results as well, the detected deviations from Gaussianity and statistical isotropy must therefore be taken to be of cosmological origin so far.

In addition to these findings, several local features including the Cold Spot could be detected with the scaling indices, which turns out to be another advantage of this method. The fact that most of them are located in the southern hemisphere confirms the conclusions concerning the asymmetries from above. Nearly all detected spots are in agreement with former analyses (e.g. \cite{2004ApJ...609...22V, 2005MNRAS.359.1583M, 2008PhRvD..78j3504P}), which confirms the existence of these local anomalies.

By accomplishing a comparison of the CMB data with surrogate maps, one focuses on the more specific assumption of random and uncorrelated phases, which is part of the Gaussian hypothesis. In addition, this method offers the possibility of a scale-dependent analysis. The scaling indices are the first measure which is used in combination with this surrogates approach. For an analysis of the 5- and 7-year observations of the WMAP satellite, the results are as follows:

Highly significant non-Gaussianities could be detected, again by performing an analysis of rotated hemispheres, for the very large scales and for the $\ell$-interval covering the first peak in the power spectrum. The results show the most significant evidence of non-Gaussianity in the CMB to date, and disagree strongly with predictions of isotropic cosmologies in single field slow roll inflation. Several checks on systematics were performed, which lead to the conclusion that the findings are of cosmological origin.

For smaller scales (i.e. higher $\ell$-ranges), it turns out that phase correlations can be easily induced by the ILC map making procedure, so that it is difficult to disentangle possible intrinsic anomalies from effects induced by the preprocessing of the data. In this case, more tests are required to further determine of these high-$\ell$ anomalies.

The SIM is the only measure in CMB analysis that was used in combination with the surrogates technique so far. Further studies that combine the surrogates method with different measures, as for example Minkowski functionals, could support these investigations and produce even more reliable results. In addition, the upcoming data of the PLANCK satellite offers an independent measurement of the CMB, and will allow investigations concerning higher $\ell$-bands.


\section*{Acknowledgments}

Many of the results in this paper have been derived using the HEALPix \citep{2005ApJ...622..759G} software and analysis package. We acknowledge use of the Legacy Archive for Microwave Background Data Analysis (LAMBDA). Support for LAMBDA is provided by the NASA Office of Space Science.


\bibliographystyle{nature}
\bibliography{/Users/gre/Desktop/EigenePapers/Literatur}

\label{lastpage}

\end{document}